\documentclass{article}

\usepackage[utf8]{inputenc} 
\usepackage[T1]{fontenc}    
\usepackage[hyperfootnotes=false]{hyperref}
\usepackage{url}            
\usepackage{booktabs}       
\usepackage{amsfonts}       
\usepackage{nicefrac}       
\usepackage{microtype}      
\usepackage[swedish,ngerman,british,american]{babel} 
\usepackage{natbib}
\usepackage{chapterbib}
\usepackage{threeparttablex}
\usepackage{multirow}
\usepackage{dpfloat, booktabs}
\usepackage{floatrow}
\usepackage{graphicx}
\usepackage{longtable}
\usepackage{amsfonts,amssymb,amsthm,epsfig,epstopdf,titling,url,array}
\usepackage{bm}
\usepackage[fleqn]{amsmath}
\usepackage{mathtools}
\usepackage{amsthm}
\usepackage{wasysym}

\def\figref#1{figure~\ref{#1}}

\def\eqref#1{equation~\ref{#1}}

\def\1{\bm{1}}

\def\vc{{\bm{c}}}

\def\vh{{\bm{h}}}

\def\vk{{\bm{k}}}

\def\vu{{\bm{u}}}

\def\vx{{\bm{x}}}

\def\vz{{\bm{z}}}

\def\mB{{\bm{B}}}

\def\mW{{\bm{W}}}
\def\mX{{\bm{X}}}

\DeclareMathAlphabet{\mathsfit}{\encodingdefault}{\sfdefault}{m}{sl}
\SetMathAlphabet{\mathsfit}{bold}{\encodingdefault}{\sfdefault}{bx}{n}

\newcommand{\R}{\mathbb{R}}

\usepackage{arxiv}

\title{Using Artificial Intelligence to Recapture Norms: Did \#metoo change
	gender norms in Sweden?}

\author{
	Sara Moricz\thanks{Department of Economics, Lund University. Email: sara.moricz@nek.lu.se Homepage: https://sara.moricz.se/
		Address: Box 7082 Lund, Sweden.} \\
	Department of Economics\\
	Lund University\\
	Lund, Sweden \\
	\texttt{sara.moricz@nek.lu.se} \\
}

\begin{document}
	\maketitle
	
	\begin{abstract}
		Norms are challenging to define and measure, but this paper takes advantage of text data and the recent development in machine learning to create an encompassing measure of norms. An LSTM neural network is trained to detect gendered language. The network functions as a tool to create a measure on how gender norms changes in relation to the Metoo movement on Swedish Twitter. This paper shows that gender norms on average are less salient half a year after the date of the first appearance of the hashtag \#Metoo. Previous literature suggests that gender norms change over generations, but the current result suggests that norms can change in the short run.
	\end{abstract}

		
\setcounter{section}{0} 
\renewcommand{\thesection}{\arabic{section}} 
\setcounter{subsection}{0} 
\setcounter{footnote}{0}
\setcounter{equation}{0}
\setcounter{figure}{0} 
\renewcommand{\thefigure}{\arabic{figure}} 
\setcounter{table}{0} 
\renewcommand{\thetable}{\arabic{table}}

\section{Introduction}
Norms are informal institutions thought to be passed down from parents to children through generations. But when do the children stop following the ways of their parents and start changing the norms? The Metoo movement might provide one such example, and maybe for the first time in history has such a popular movement left an easily accessible trace as more and more private conversations take place online on social media. This article contributes to institutional economics, firstly, by providing a method for automatically detecting norms in text data. Then, the article shows that gender norms as reflected on Swedish Twitter are less adhered to the period after the Metoo event.

Previous research conceptualizes norms as long-run phenomena and authors present evidence of gender norms being constant over generations \citep{alesina2013origins,fernandez2004mothers,nollenberger2016math,williamson2000new,bisin2001economics}. At the same time, gender norms seem to be important: gender norms explain a part of the gender gap in mathematical test scores \citep{pope2010geographic,fryer2010empirical,guiso2008culture,nollenberger2016math,spencer1999stereotype} and indications exist of them also affecting the gender gap in the propensity to negotiate \citep{leibbrandt2014women,bowles2007social,mazei2015meta}.\footnote{Some indication also exists of various personality traits, foremost agreeableness, being differently rewarded on the labour market dependent on if it is women or man exhibiting it \citep{mueller2006estimating}.} Gender norms are hard to measure, which limits the research field. \citet{bertrand2015gender} provide the most recent contribution in the measurement of gender norms. They ingeniously measure the norm “a man should earn more than the women in a couple” directly from observational data. The usage of observational data overcomes social desirability bias: that participants in surveys or experiments act according to what they think is acceptable behavior. Unfortunately, the type of observational data they use only measures hard life choices, such as educational decisions and marriages. One major contribution of this article is to use observational data while comprehensively measuring gender norms. The comprehensive definition of gender norms includes the myriads of aspects defining the gender roles women and men face in their everyday life. The other major contribution of this article is to show that gender norms are less followed in Swedish tweets in the wake of the Metoo event, which complement existing studies by suggesting that gender norms can change in the short run.

Norms are informal institutions, and North defines institutions to be “the humanly devised constraints that shape human interaction” \citeyearpar[p. 3]{north1990institutions}. \citet{akerlof2000economics} brought norms into the economic utility framework: In their identity framework, people belong to different social categories with different prescriptions for how they should behave. This paper postulates the following data generating process: unobserved gender norms giving rise to observed gendered language, i.e. males and females are being depicted differently in texts. To obtain a tool for automatically detecting gender norms, I censor Swedish tweets by replacing all instances of the words \textit{he} and \textit{she} with a blank spot, \_\_\_, and train a neural network algorithm to fill in the correct word. Neural networks are overarching machine learning algorithms which allow analysts to do broad searches for good models. They are state-of-the-art methods in natural language processing to translate between languages and capture the sentiment of a text. A person asked to fill in a blank spot with \textit{he} or \textit{she} would use previous knowledge of gender prescriptions to make a good guess. Instead, with my data setup, the training yields a model defining the gender prescriptions as the average of Twitter users’ differential portrayals of males and females in the training tweets. 

The current method entails an objective definition of gender norms. One dimension of gendered language (\textit{he}/\textit{she}) is used to infer the rest of the dimensions. The objective approach is important insofar that analysts can hold norms implicitly affecting their choices regarding what they count as a subnorm. \citet{BolukbasietAl2016} create an algorithm to remove unwanted gendered words in word vectors, which necessitates that an analyst provides a list of unwanted words. Word vectors are vector representation of words obtained through clustering, also used in this paper. \citet{wu2017gendered} finds that words related to physical appearance are high-predictors for posts on the Economics Job Market Rumors forum and interprets such words as indicating the presence of gender norms. She uses a penalized logistic regression with the counts of words as the independent variables, which, again, is a model-technical building block in this paper. The main difference between the two other articles and this one is in the end goal: This article strives to encapsulate norms into a model to be used as evaluation tool on a new sample in a research design.

I have downloaded all Swedish tweets from Twitters API, \textit{he} and \textit{she} are frequent words and the total he/she-sample size is large at 2 million tweets. Twitter data is written directly by people without an editor and norms are therefore more likely to be reflected in tweets than in, for example, newspapers. This paper contrasts an LSTM neural network algorithm to a baseline Naive Bayes algorithm. The main difference between them is that the neural network model has the ability to learn phrases, i.e. it takes word order into account, whereas the Naive Bayes model does not. In the main version of the models are highly-gendered words, such as \textit{mum} and \textit{dad}, censored to allow for a more nuanced definition of gender norms. The models are trained on 800 000 he/she-tweets. On a new set of 200 000 out-of-sample tweets the neural network model predicts 77 percent correct whereas the Naive Bayes model predicts 75 percent correct. Words related to sport are high predictors for males and words related to personal relationships are high predictors for females. Then, this model is used as a tool to measure the change in gender norms followed in relation to the Metoo movement.

The Metoo movement started when an American actress asked other women to report misogynistic behavior. The hashtag peaked on Swedish Twitter on 17th of October 2017 and did not exist before that. The Metoo event provides a good context for evaluating if gender norms change rapidly, the exact date for its occurrence is exogenous while it spurred a collective response in the Swedish society. The movement lead to women sharing their experience of sexual harassment and it is likely that it leads to a norm shift towards more awareness and less acceptance of the sexualization of women. The movement also spurred an intensive public debate on gender issues and the norm shift may equally well include less acceptance towards stereotyping the genders in general. The current encompassing measure capture the possible norm shift. The model is used to investigate how much gender norms are followed by measuring its predictive accuracy, i.e. how often it predicts the correct word \textit{he} or \textit{she} in the blank spot in tweets. For example, let us assume that the model have encoded “struggle”, “math” and “homework” as following a female gender stereotype and before the event tweets such as “my daughter she struggles with her math homework” are prevalent, then the model has high predictive accuracy indicating that gender norms are followed. If after metoo many daughters are called “competing for good grades” instead and this is encoded as indicating a male gender stereotype, the model would detect the change by a decrease in its predictive accuracy.

The research design uses an additional 1 million out-of-sample he/she-tweets from May 2017 to May 2018, data from five months leading up to the event to six months after, to evaluate the Metoo movement. The model is trained on data one year before, from May 2016 to May 2017, which defined gender norms to the average of this time period. Tweets that contain the hashtag \#Metoo and correlated hashtags are filtered out not to capture the effect of a more intensive gender debate. The Metoo movement is colinear with time and the measured effect always runs the risk of being driven by a third factor. To decrease such possibilities, the research design is such that the 200 000 out-of-samples tweets mentioned before (used to test model performance) match the year before and can be used as a comparison group. In my preferred specification, those tweets allow controlling flexibly for time trends by time fixed effects. The tweets include user identification numbers which allow for user fixed effects on the 35 000 users that tweets at least once on either side of the Metoo event as a robustness check. Before the Metoo event gender norms are on average more followed in tweets than the year before, but after the Metoo event my preferred estimate shows that norms on average are followed 1.9 percent less over the sample mean. The gender norms followed before the event on Swedish Twitter would be erased by 18 Metoo movements. The comparison is not normative since the data-driven definition of gender norms does not include judgment.

Swedish twitter users are younger and slightly more male than the general population \citep{ISS2016} rendering the norms and the effect on this demographic to be estimated. At the same time, younger persons are likely to have got a stronger experience of the Metoo event as it was largely an online movement. The result should be interpreted in the broader context of changing norms in the short run. The Metoo movement is an example of the newer kind of popular movement spread via social media, a form of bottom-up revolution. The current result of changes in the following of gender norms, therefore, points to online mass population movements affecting an overall norm in the short run. For the women and men participating in such movements, it is important to see that they do bring change.

The rest of the paper is organized as follows: Section 2 reviews articles investigating gender norms. Section 3 presents the proposed method for measuring norms and discusses how it differs compared to previous measures. Section 4 presents the data. Section 5 illustrates the neural network model,  whereas appendix A presents model selections and specifications in closer detail.  Section 6 presents model results. Section 7 describes the Metoo event in Sweden, the research design behind the evaluation of the Metoo event and the results on how metoo have impacted gender norms. Section 8 concludes.

\section{Previous literature}

Norms are informal institutions. North define institutions to be “the humanly devised constraints that shape human interaction” \citeyearpar[p. 3]{north1990institutions}. Norms are informal since other people punish a person for breaking them, rather than a formal third-party, such as the police. In \citet{akerlof2000economics}'s identity framework people belong to different social categories defined by prescriptions for how they should behave. Breaking the prescriptions affect one's own and other's utility, essentially imposing an externality, leading to people being punished for following a specific identity.  

Norms are generally thought to be changing slowly, over the course of centuries or millennia \citep{williamson2000new}. \citet{bisin2001economics} formalize theoretically how parents and the surrounding environment transmit norms to children. \citet{alesina2013origins} show that gender roles transmit over long time periods; immigrant children of parents from societies in which the plow historically was used (a proxy for a norm of males working outside the household) think that a woman's place is in the household to a greater extent. \citet{fernandez2004mothers} provide evidence of gender attitudes transferring within families; wives of men whose mothers had been working during WW2 is more likely to participate in the American labour force.\footnote{ \citet{nollenberger2016math} corroborates transfers of attitudes within families.} At the same time, norms can change quickly by external shocks. \citet{fortin2015gender} shows how the evolution of gender norms and female labour force participation co-varies over time. She argues that the AIDS scare changed gender norms attitudes explaining parts of the slow- down of the increasing trend in female labour force participation from the mid-1990:s in the US. \citet{goldin2006quiet} argues that young women in the 1960:s anticipated a life on the labour market giving rise to a “revolutionary phase” observed in a multitude of variables from the late 1970:s in the US. In this article, the newer method and the exogenous Metoo event makes it possible to investigate if gender norms can change in a shorter timespan.  

\begin{table}[H]
	\begin{centering}
		\resizebox{1\textwidth}{!}{
			\begin{threeparttable}
				\begin{tabular}{lll}
					\hline
					Paper                  & Norm                                   & Measured by                                          \\ \hline 
					Bertrand et al. (2015) & "a man should earn more than his wife" & relative income between spouses                      \\
					Wu (2018)              & "the gender norm among economists"     & gendered language in the Economics Job Market        \\
					&                                        & Rumors forum                                         \\
					Alesina et al. (2013)  & "women should stay in the household"   & historical plough usage, female labour force partic- \\
					&                                        & ipation, share of female firm owners, proportion of  \\
					&                                        & women in parliament, two World                       \\
					&                                        & Values Survey questions*                             \\
					Fortin (2015)          & "women should stay in the household"   & index variables created from the surveys General     \\
					&                                        & Social Survey (US) and National Longitudinal Sur-    \\
					&                                        & vey (US)*                                            \\ \hline
				\end{tabular}
				\begin{tablenotes}[flushleft]
					\item[] Note: {*} a list of the survey questions is found in appendix table \ref{tab:Survey questions used to capture gender norms}.
				\end{tablenotes}
			\end{threeparttable} 
		}
	\end{centering}	
	\caption{\label{tab:Papers which measures norms directly} Papers which measures norms directly}	
\end{table}

In the last twenty years, norms have been recognized as important in explaining economic phenomena, but the lack of available data on norms most likely hampers the research field in Economics. To my knowledge, \citet{bertrand2015gender} is the first paper which investigates a gender norm directly from observational data. More specifically, they investigate how the norm "a man should earn more than his wife" affects a multitude of variables such as sorting into couples and labour market participation. Strikingly, they show that the gender gap in non-market work increases with a rising share of female family income. Also, \citet{wu2017gendered} uses observational data, in this case, blog post from the Economics Job Market Rumors forum. She finds that words related to physical appearance frequently appear in posts related to females. A need for more observational data on norms, in general, exists to further the field. Observational data overcome social desirability bias, that participants in a survey or an experiment change their answer or behavior due to being surveyed. In addition, it is difficult to obtain survey data: the lower half of Table \ref{tab:Papers which measures norms directly} lists the two papers pertaining to gender norms reviewed which measure gender norms directly from surveys.

Many papers are reactionary in their research design; researchers try to disprove that various observed differences between men and women are due to biology. One observed difference between men and women is the gender gap in mathematical test scores, where boys usually achieve better. The gap varies over states/countries and is positively correlated with measures of gender inequality \citep{pope2010geographic,fryer2010empirical,guiso2008culture}. Biological differences are unlikely to vary over states/countries and the pattern points to the importance of gender norms. \citet{nollenberger2016math} show that the gender gap decreases in second-generation immigrant children if their parents' source country is more gender equal. Their research design rules out possible confounding state/country-factors, thereby strengthening the interpretation. Alike research designs are inventive, but the outcome variable becomes the gender dimension investigated, for instance mathematical test scores, which unfortunately is quite limited in scope.

\section{Presentation of my method and how it compares to previous measures of norms}

\begin{table}[H]
	\begin{centering}
		\noindent\resizebox{1\textwidth}{!}{
			\begin{threeparttable}
				\begin{tabular}{@{}lc@{}}
					\toprule
					Tweets                                                                                                                                                                                       & Predicted Class \\
					\midrule
					& \\
					\textbf{English}                                                                                                                                                                                      &                 \\ \hline
					\multirow{2}{.85\textwidth}{(1) \textgreater \textgreater ... Don't know which reality \_\_\_ lives in. *Smiling Face With Open Mouth and Cold Sweat Emoji*}                                                                  & She             \\
					& \\
					\multirow{2}{.85\textwidth}{(2) \textless{}user\textgreater{} I don't like Pippi either. Hate how \_\_\_ always buys \textless{}female friends/male friends\textgreater{} and all prudence.}                                      & She             \\
					& \\
					\multirow{2}{.85\textwidth}{(3) \textless{}user\textgreater{} \textless{}user\textgreater{} Swedish nazi brags about that \_\_\_ has killed in Ukraine white Europe is a utopia common among SD:s … \textless{}url\textgreater{}} & He              \\
					& \\
					\multirow{1}{.85\textwidth}{(4) What was it \_\_\_ was called at FIFA? Infantilo?}                                                                                                                                           & He              \\
					& \\
					\textbf{Swedish}  &                 \\ \hline
					\multirow{2}{.85\textwidth}{(1) \textgreater \textgreater ... Vet inte vilken verklighet \_\_\_ lever i. *Smiling Face With Open Mouth and Cold Sweat Emoji*}                                                                 & Hon             \\
					& \\
					\multirow{2}{.85\textwidth}{(2) \textless{}user\textgreater{} jag gillar inte heller Pippi. Avskyr hur \_\_\_ ständigt köper sina \textless{}väninnor/ vänner\textgreater{} och all präktighet.}                                    & Hon             \\
					& \\
					\multirow{2}{.85\textwidth}{(3) \textless{}user\textgreater{} \textless{}user\textgreater{} Svensk nazist skryter om att \_\_\_ har dödat i Ukraina vitt Europa är en utopi vanliga SD:are … \textless{}url\textgreater{}}        & Han             \\
					& \\
					\multirow{1}{.85\textwidth}{(4) Vad var det \_\_\_ hette på FIFA? Infantilo?}                                                                                                                                                & Han             \\
					& \\ \bottomrule
				\end{tabular}
				
				\begin{tablenotes}[flushleft]
					\item[]Note: The dependent variable, if \textit{she} or \textit{he} should be placed in the blank spot, is binary. The model uses the sequence of words, the fill-in-the-blank tweet, to predict which class, he or she, it belongs to. In the table is the predicted class equivalent to the true class. Words are masked both due to protect privacy and to reduce noise, e.g. \textless{}user\textgreater{} and \textless{}url\textgreater{}. Other words are masked to take away an “obvious” gender dimension. For example, there exist a female and a male version of the word \textit{friends} in Swedish, \textit{female friends} (vänninor) and \textit{male friends} (vänner). The masking is illustrated with that \textless{}female friend/ male friend\textgreater{} is considered a word instead of the original two. The translation from Swedish is made by the author.
				\end{tablenotes}
			\end{threeparttable}
		}
	\end{centering}	
	\caption{\label{tab:examples}Examples of Input Tweets}	
\end{table}

This paper assumes the following the data generating process: There
exists an unobserved gender norm (or at least it is hard to quantify
what it consists of and how strong it is) giving rise to observed
gendered language. Gendered language is that people write differently
depending on if they depict a male or female. As an illustrative example,
a nurse of female sex would be referred to as a \textit{nurse}, whereas
a nurse of male sex would be referred to as a \textit{male nurse}.
The method used in this paper consists of training a model to detect
gendered language. Table \ref{tab:examples} illustrates it: The model
learns to put in the correct word, \textit{he} or \textit{she}, in
tweets where blank spots have replaced those two words. A binary he/she-
variable, made up of the correct words, is used as a dependent variable
to infer gendered language. The independent variables are the sequence
of all words in the training dataset vocabulary, including the blank
spot. 

A person that would guess if the tweets in Table \ref{tab:examples} should be filled with \textit{he} or \textit{she} would use previous knowledge of gender prescriptions to make a good guess. The model does the same but uses previous knowledge from all the tweets it has been trained on. More specifically, the model learns the average of how Twitter users choose to depict women and men in separate ways in the training data. It learns to separate because the training setup renders the model to be discriminatory. For instance, it can only use words and phrases describing \textit{he} more than \textit{she} to predict to the he-class. I argue this is equivalent to learning a definition of gender prescriptions. If there were no gender norms, the Twitter users would not systematically choose to portray females and males in specific ways, and the model would not learn any words and phrases that help it in predicting \textit{he} and \textit{she}. The stronger the gender norms, the more differential depictions of the genders exists in tweets, the easier it is for the model to learn words and phrases that predict the genders. Thus, this paper defines the gender norms to be a set of boundaries of different intensity that defines “being a woman” and “being a man”, which follows North's and Akerlof and Kranton's definitions \citeyearpar{north1990institutions,akerlof2000economics}. For example, in the tweets, women are overrepresented as pictured to be involved in personal relations, such as being friends and colleagues, and the model learns that this defines "being a women". The interpretation becomes that the social punishment for not being involved in such personal relations is higher for a woman than a man. 

Persons can on a qualitative basis observe the gender norms, but to quantify the change of them are more difficult. A model capturing a definition of gender norms can be used to measure change over different samples, both by investigating sequences of words used and the overall predictive accuracy. This paper uses predictive accuracy, i.e. how many he- and she-tweets the model predicts correct, to measure if a new sample of tweets reflects the already-defined norms. The measure interprets as a reflection of to which extent third-party persons will experience the past gender norms by reading the new tweets. Another possible interpretation of the measure is that it reflects the average following of the past gender norms by the users posting the new tweets. If a user chooses not to tweet due to a change in the following of the past norms the missing tweet does contribute to the norm change by its absence, i.e. by nobody experiences the tweet. According to \citet{north1990institutions}, the humanly devised constraints (the norms) are shaped by human interaction (the posting and the reading of tweets). In sum, the measure indicates a change of norms.

The proposed method will yield a much more encompassing definition of norms than previously employed. It uses all the myriads of aspects that Twitter users choose to write to depict the genders in separate ways to create the definition. With the proposed method the analyst only selects one group feature to infer the rest, whereas, with surveys, experiments or research designs the researcher decides which sub-norm to investigate. I expect the data-driven method to take into account male gender roles to a much larger extent than what has previously been done in research. For example, by using surveys, researchers ask respondents questions about how they feel about women working in the labour market, but usually do not ask questions about males taking a larger responsibility for working in the household (see appendix table \ref{tab:Survey questions used to capture gender norms}). Also, the proposed method will yield an objective definition of gender norms. In this paper, one dimension of gendered language,\textit{ he} and \textit{she}, is used as an independent variable to infer the rest of the gendered language- dimensions. An alternative way of quantifying gendered language would be to let persons mark tweets as gendered. Using such training data would teach the model to capture an average of the marking persons' gender stereotypes. Since persons might hold norms subconsciously, the marking might be biased. The current method not only allows for the usage of text data, it also lets the data decide what the norm consists of, yielding an encompassing and objective definition.

\section{Data}

I have downloaded 100 million tweets marked with being in Swedish from May 2016 until May 2018 from Twitter's API. Twitter data is published directly by people without an editor. Since people write messages to each other that is more closely related to how they speak in real life, it is more likely that tweets capture gender norms than, for example, books and news articles. The data used do not include retweeted tweets.  The main dataset consists of tweets which include the words \textit{he} (han) or \textit{she} (hon). I replace the true word indicating gender (\textit{he}, \textit{she}) with a placeholder and the true word becomes a variable in the dataset. It is ambivalent which gender a tweets containing both \textit{he} and \textit{she} is about. Such tweets are not included in the dataset as they only constitute 0.7 percent of tweets containing any of the two words. Twitter users are anonymous for all practical purposes, for example, there is no information on if it is a man or a woman who writes a specific tweet.\footnote{The name chosen by a twitter user is given in the data and might indicate the gender of the user. However, since there is no easy way of judging the measurement error by approximating the user's gender from the twitter name I refrain from using the information. In general, see \citet[p.83-84]{steinert_threlkeld_2018} for a discussion on common background variables that does not exist on Twitter users.}

\begin{table}[H]
	\begin{centering}
		\resizebox{0.8\textwidth}{!}{
			\begin{tabular}{lllccc}
				\hline
				\multirow{3}{*}{Set} & \multirow{3}{*}{Function} & \multirow{3}{*}{Time} & \multicolumn{3}{c}{Count of tweets}                                              \\ \cline{4-6}
				&                           &                       & \multirow{2}{*}{Original} & \multirow{2}{*}{10-25 words} & Hashtags uncorrelated \\
				&                           &                       &                           &                              & to \#metoo            \\ \hline
				Training             & Estimate parameters       & Year 1                & 984 337                   & 676 466                      &                       \\
				Validation           & Model selection           & Year 1                & 246 085                   & 169 184                      &                       \\
				Test                 & Evaluate model            & Year 1                & 307 606                   & 211 242                      & 203 691               \\
				Evaluation           & Evaluate model            & Year 2                & 1 396 705                 & 1 017 414                    & 989 028               \\ \hline
		\end{tabular} }
		\caption{Sets and sample sizes}
		\label{tab:Sets-and-sample}
	\end{centering}
\end{table}

From the he/she-tweets several subsets are formed. I create two main sets; Year 1 at 1.5 million tweets spanning 2016-05-13 to 2017-04-30 and Year 2 at 1.4 million tweets spanning 2017-05-01 to 2018-04-30. The Metoo event takes place in the middle of Year 2. Table \ref{tab:Sets-and-sample} displays the sets, their functions and their sample sizes. The he/she- tweets of Year 1 are further subdivided into a training, validation and test set according to a random 64-16-20 percent split. The subdivision to a training, validation and test set is standard in the machine learning paradigm. The training set is used to estimate parameters. The validation set is used for model selection, i.e. to estimate hyperparameters. The test set is reserved for evaluating the chosen model on unseen data. The tweets of Year 2 is also a test set in the machine learning- sense, but as it performs a separate function in the evaluation of the Metoo movement I make it distinct and call it the evaluation set. Because tweets with very different word counts are difficult to handle in the same model, the sets are subsampled only to contain tweets that include 10-25 words. The removal cuts the sample sizes to about 70 percent of the original. An alternative would be to train various submodels, but as the main goal is to capture gendered language the current solution is deemed sufficient. An additional reason for the sample cut is that tweets with a word count below 10 contain little information to predict on. The models are trained and validated on 845 000 tweets and tested on 211 000 tweets. The tweets of Year 2, the evaluation set, is used to evaluate the Metoo movement by measuring the change in gender norms vis-a-vis the test set of Year 1. The two sets are subsampled to only include tweets which do not have a hashtag correlated to the metoo-hashtag and by removing missing dates.\footnote{Due to downloading issues data are partly missing for 22 dates from Year 1 and 44 dates for Year 2 as displayed in appendix table \ref{tab:Missing-UTC-dates}.} The final sample size for assessing the effect of the Metoo movement is 1.2 million tweets.

A model will learn that names, such that \textit{Erik} and\textit{Anna}, refer to males and females. A definition of gender norms is not nuanced if the model simply learns such names. Thus, in the main version of the model all first names are censored. The name list is retrieved from Statistics Sweden \citep{scb2016females,scb2016males}. It contains all first names used by any Swedish residence and the number of female and male name holders. 

\section{Model illustration}

An LSTM neural network model is presented in this section as it outperforms other models, as shown in Appendix A. The appendix presents the model specifications and selection in much greater detail with the terminology used in the neural network- literature. 

\subsection{Overview of machine learning concepts}

Machine learning techniques have recently found increasing applications in economics \citep[p. 99-103]{mullainathan2017machine}. The methods are characterized by the functional form being learned from the data. Although papers within economics use machine learning methods, mostly papers in the finance literature have adopted neural networks. A neural network is not a model, it is an algorithm in the sense that the analyst sets up limits and a good model is searched for within those.\footnote{However, the convention of referring to neural networks as models seems established and consequently I refer to neural networks as models in this paper.} Neural networks are universal function approximators \citep{hornik1991approximation}: theoretically, they can approximate any functional form and therefore can the algorithm find any model. The result is based on the network searching in such broad limits that it is computationally infeasible and, in practice, the analyst creates boundaries within which the network chooses the best fit. In general, networks are used for high-dimensional data and designed to contain many other types of machine learning algorithms. 

A common task is to predict $Y$ from the data, $D=[Y,X]$. More precisely,
the analyst forms an error function (loss, objective, cost function):

\begin{equation} \label{eq:1} 
E=-P(Y_i,F_c(X_i,A,\Omega,B))
\end{equation}

where $P()$ is a performance measure of the model as a function of the dependent variable $Y$ and the candidate model $F_c(X_i,A,\Omega,B)$. The error function (\ref{eq:1}) behind a neural network can include many different specifications of $P$ and $F_c()$, i.e. networks can be differently designed. I introduce the concept of having a candidate model, $F_c()$, to be able to compare machine learning versus more standard statistical approaches. The reason for calling it candidate model is that the specification also includes regularization, $\Omega$, various features existing to penalize overfitting (overtraining/ model complexity), and adherent hyperparameters, $A$, that guide model selection. The candidate model is not akin to the model obtained by training the network. The functional form of the obtained model is learned by splitting the original data in two parts: a validation set used for model selection and a training set used for estimating model parameters. 
By choosing the $A$ hyperparameters that give the lowest error in eq. (\ref{eq:1}) on the validation set a model is selected that is not overfitted, i.e. which have good generalization performance. The model parameters, $B$, are estimated on the training set for each hyperparameter combination. By estimating hyperparameters on a new sample, the validation set, new error terms are drawn, and data not used to estimate its parameters evaluates the candidate model.\footnote{In the above framework, the lasso regression seems to be most commonly used in economics and is therefore examplified. For the general error function in eq. (\ref{eq:1}) to be equivalent to a lasso regression the analyst selects the performance measure to be the mean square error, $P= \sum_{n=1}^{i} (Y_i-F_c(X_i,A,\Omega,B))^2$. Also, the analyst decides that the candidate model is $F_c(X_i,A,\Omega,B)= X_i^Tb + a\sum_{k=1}^{k} \mid b_k \mid$. Thus, the analyst decides to investigate the set of linear functions (first part) and to reguralize the model by shrinking large coefficients with the L1 norm (second part).} In most of standard statistics\footnote{The major exception is in time series analysis when the forecast error on out-of-sample data decides model selection, which is equivalent to minimizing the performance measure on the validation dataset.} the training-validation set split is absent, and the data analyst proposes a candidate model $F_c()$. Residuals from the same data approximate the error terms. On a conceptual level, in most of standard statistics the analysts are more involved in the algorithmic process of finding a good model (especially if an analyst tests a model proposed by another on new data), whereas in machine learning an automated algorithm performs more of the task. 

Verification of the model is achieved by evaluating the learned model on unseen data, a test set, separate from the training- and validation set mentioned previously. Machine learning methods generally and neural networks specifically are “black box” approaches. “The black box” exists since networks are designed to handle high-dimensional problems and consequently it is hard for the analyst to interpret the parameters of the learned model. Neural networks tend to be used to solve predictive tasks; they are congruent with inference on the model level.

\subsection{Model specification}
The LSTM neural network model specifies as follows: 

\begin{equation*}
	P = \sum_{n=1}^{1} Y_i \log F_c(X_i,A,\Omega,B)+(1-Y_i)\log (1-F_c(X_i,A,\Omega,B))
\end{equation*}

where $Y_i$ is a binary indicator on if \textit{he} or \textit{she} is written in the blank spot. The performance measure $P$ is the cross-entropy error. $P$ is in itself equivalent to optimizing the log-likelihood function behind a Logit model. The candidate model $F_c(X_i,A,\Omega,B)$ is a large parametric model which will search a large scope of possible functional forms to learn:

\begin{multline*} 
	F_c(X_i,A,\Omega,B_i)= f_o(\sum_{m=1}^{M}b_m^of_h(h_{m,t=T})+c^h) \\
	f_o(z) = \sigma(z) \\
	f_h(z) = \left \{   
	\begin{tabular}{ll}   
		z & if z$>$1  \\   
		0 & if z$<$1 \\
	\end{tabular}
	\right . \\
\end{multline*}

Regularization in the designed network enters by the hyperparameter the number of hidden nodes (M), which decides the number of $b_m^o$ coefficents and $h_{m,t=T}$-processes. The number of hidden nodes generally guides how large a neural network model should be and thus, how well the model can fit the data.\footnote{Regularization enters in three ways but only one is explicitly shown in the specification. The other two hyperparameters are the following: The second hyperparameter is training time because the error function is numerically optimized by updating the candidate model with a number of randomly selected observations at each round (stochastic gradient descent). The third hyperparameter is the percentage of $b_m^o$:s present for each individual update, which pertains to a computational efficient model averaging technique (dropout). } The candidate model is a panel-data model as each $h_{m,t=T}$-process is specified by:

\begin{align*}
	h_{m,t} = &\tanh(c_{m,t})\sigma(x_{t,i}B_m^{u1}+h_{m,t-1}B_m^{u2}+c_m^u) \\
	c_{m,t} = &c_{m,t-1}\sigma(x_{t,i}B_m^{f1}+h_{m,t-1}B_m^{f2}+c_m^f)+
	\tanh(x_{t,i}B_m^{c1}+ \\
	&h_{m,t-1}B_m^{c2}+c_m^c)\sigma(x_{t,i}B_m^{p1}+h_{m,t-1}B_m^{p2}+c_m^p) \\
\end{align*}

The presentation of the specification is brief, the above equation is a short form of a LSTM node which is explained in appendix A, subsection "(3) LSTM". The main take-away message of the current presentation is that the candidate model is a large parametric model which will be reduced during training. 

The inputs $x_{t,i}$ to the network are word vectors. The usage of word vectors is today common within natural language processing. The word vectors are parameters to the network which already are specified, $x_{t,i}= d_{t,i}B^r$. Dummy-variable encoded vectors, $d_{t,i}$, representing the \textit{t}:th word in the \textit{i:}th tweet, get reduced by the $B^r$ matrix to word vectors $x_{t,i}$. Each word vector is a k-dimensional vector representing one word and each word vector element is a parameter "looked up" in the $B^r$ matrix.

\subsection{Word vectors}
I choose to train my own word vectors by the Skip-Gram method of \citet{mikolov2013efficient}. Available Swedish word vectors are based on Wikipedia text and as the type of language used differs a lot to Twitter, using my own-encoded vectors likely enhances the performance of the model. 

\begin{figure}[H]
	\begin{centering}
		\includegraphics[width=1\textwidth]{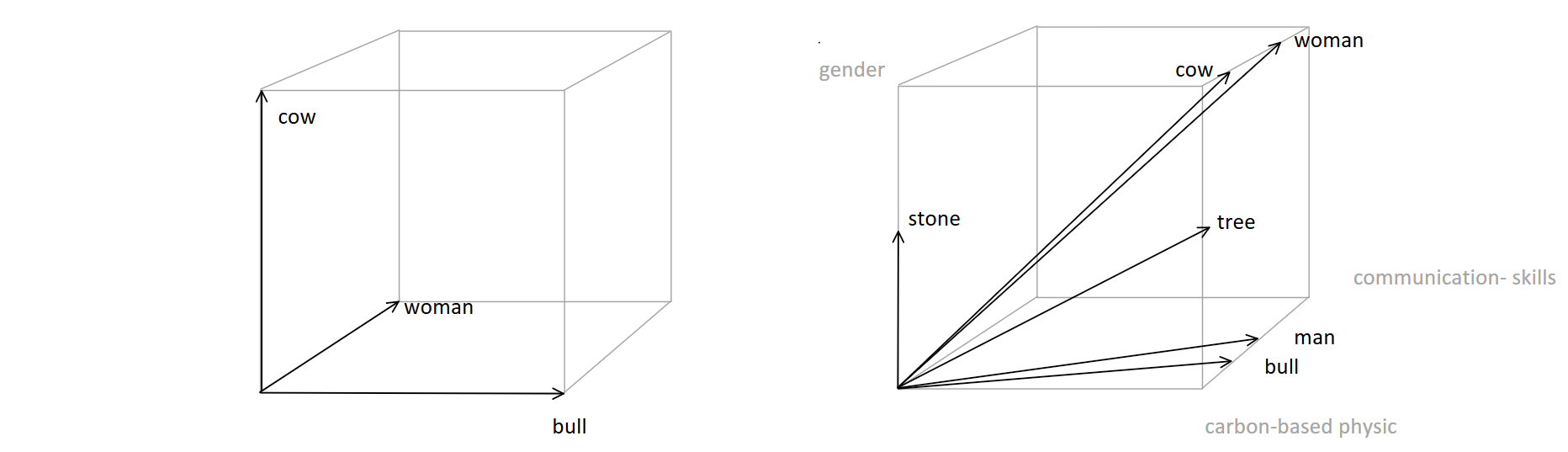}
		\par\end{centering}
	\centering{}\caption{\label{fig:Example-of-categorical}Example of dummy variable representation (left) and word vector representation (right)}
\end{figure}

An intuitive explanation of the dimensionality reduction performed in the first part of the model, going from dummy-variable encoded vectors to word vectors follows. Words are usually thought of as dummy-variables, but it might not be a good representation. Let us take \textit{cow}, \textit{women} and \textit{bull} as an example. Formulated as being three dummy variables would mean that they are orthogonal to each other, as visualized to the left in Figure \ref{fig:Example-of-categorical}. This is clearly not the case as \textit{cow} and \textit{women} are both of female gender and they have some relation that is not captured with a dummy-variable encoding. Words can be encoded into many different vector spaces, without any inherent truth about whether it provides a good representation or not. One example of such a vector space is the right one in Figure \ref{fig:Example-of-categorical}, which illustrates the encoding of words to word vectors developed by \citet{mikolov2013efficient}. Such word vectors have good properties, for example, it is possible to perform vector algebra on them and get sensible results, which is illustrated in the figure.\footnote{Various paper exists evaluating the properties of word vectors. For example, \citet{arora2016linear} investigate polyesmious words, words where the letters are the same but where the meaning differ, and find that they are placed in the center of their respective meanings.} When using the pretrained word vectors, this paper uses a representation, again with no inherent truth of whether or not it is good. If it is a good representation or not is judged by how good the obtained model is at predicting on the test set.

In a related paper, \citet{BolukbasietAl2016} classify certain words as gendered in an unwanted way and show that this dimension exists in the word vectors trained on English Google News. They provide an algorithm to remove the unwanted gender dimension in the word vectors for usage in different downstream applications. \citet{BolukbasietAl2016} find that one dimension is sufficient to detect unwanted gendered words in the word vector space. If this pattern generalizes to gendered words in general, it suggests that a weighted sum of word vector elements would be a good representation for the gendered-words intensity of a sentence. Appendix A shows that performance increases with the non-linear LSTM model, showing that additional predictability over a weighted sum exist for the present case. The main difference between the papers is that this paper detects gendered language in sentences, instead of removing predefined unwanted gendered words from word vectors.

\subsection{Masking of Words}

A model will use words such as \textit{boy - girl} and \textit{Anna - Erik} to predict if a tweet is about a man or woman. One could argue that a good definition of gender norms ought to be more nuanced than capturing such “too obvious” gender-colored language. Thus, this paper also trains the models on masked data where "too obvious" gendered colored words are censored. As an example, in the masked data the fictive word \textit{boygirl} replaces the original words \textit{boy} and \textit{girl}.  Both raw/unmasked and masked versions are trained, but when evaluating the Metoo movement the masked version is used. The mask version of the main model uses three percent fewer words than the unmasked version. 

The words used for masking divides into three groups; pairs of male-female words, family names of famous persons and first names. I have selected words to mask in the following way: Firstly, I selected pairs of male-female words (\textit{boy - girl}) from lists of high predicting words from the Neural Network or Naive Bayes model. I extended the list with the words' respective neighbors according to the pretrained word vectors (\textit{boy's - girl's, the boy - the girl}). I deemed some high predicting words to be impossible to map, such as \textit{witch} and \textit{tinder dude}, and such words are not masked. Secondly, I selected family names of famous persons from the same lists. Lastly, I selected all Swedish resident's first names, except of names that also is another word in Swedish (\textit{kalla, du}) or English (\textit{star, honey}), from the name list obtained from Statistic Sweden.\footnote{For additional details please see appendix A.} For the two later groups, the fictive word \textit{familyname} and \textit{firstname} replaces all names. See the table "Words used for masking" in the online appendix for the lists.

\section{Model result}

Table \ref{tab:main-results} evaluates the models on the test set of Year 1. Initial model selection in Appendix A shows that the best-performing model is a one-layer neural network LSTM model. The table presents it alongside a Naive Bayes model, which function as a baseline. The Naive Bayes model is a well-established model for similar tasks \citep[p. 210-211]{hastie2009elements}. The main difference between them is that the neural network model takes into account the sequencing of words, i.e. can learn phrases, whereas the Naive Bayes model does not. The first row of table \ref{tab:main-results} presents the chosen performance metric, ROC AUC-scores. A ROC AUC-score of 0.5 means that the model has no predictive power: the model cannot separate the genders in text. A score of 1 means that the model has perfect predictive power: the model can completely separate the genders. The first two columns present model results for the Neural Network- and Naive Bayes models that are trained and evaluated on the raw/ unmasked data. Those two models use words such as \textit{mum} and \textit{dad} to capture gendered language. Since this might not be in line with the type of gendered language one wants to capture, the last two columns of the table present the masked versions of the models. In the masked data fictive words such as \textit{mumdad} replace gendered words such as \textit{mum} and \textit{dad }. Inspection of the ROC AUC-scores in table \ref{tab:main-results} shows that the Neural Network model always is better at predicting than the Naive Bayes model. The Neural Network-model is an overarching model type, the Naive Bayes model is a submodel of it. The differences between them are small, 0.039 and 0.026 ROC AUC-score points for the unmasked and masked version respectively. The Neural Network model allows for the usage of sequential information, that one word comes before another, and the result points to that this not is very useful in predicting the he/she-variable. The Naive Bayes model only uses the relative frequency of words in each class to predict. The result shows that genders to a large extent can be discerned in text by the words they are bundled with. The table also shows, as expected, that the scores decrease in the masked versions of the models. The masked version of the Neural Network model has a ROC AUC-score of 0.76, showing that the genders indeed still can be separated. 

\begin{table}[H]
	\begin{centering}
		\noindent\resizebox{0.8\textwidth}{!}{%
			\begin{threeparttable}
				\begin{tabular}{lccccc}
					\hline
					& \multicolumn{2}{c}{Unmasked} &  & \multicolumn{2}{c}{Masked}   \\ \cline{2-3} \cline{5-6}
					& Neural Network & Naive Bayes &  & Neural Network & Naive Bayes \\ \cline{2-3} \cline{5-6}
					&                &             &  &                &             \\
					ROC AUC           & 0.8496 & 0.8107 &  & 0.7629 & 0.7372 \\
					&        &        &  &        &        \\
					Unbalanced Sample &        &        &  &        &        \\ \cline{1-3} \cline{5-6}
					Accuracy          & 0.8242 & 0.8058 &  & 0.7748 & 0.7522 \\
					Sensitivity       & 0.4020 & 0.4015 &  & 0.2133 & 0.3708 \\
					Specificity       & 0.9656 & 0.9412 &  & 0.9629 & 0.8799 \\
					&        &        &  &        &        \\
					Balanced Sample   &        &        &  &        &        \\ \cline{1-3} \cline{5-6}
					Accuracy          & 0.7529 & 0.7224 &  & 0.6850 & 0.6619 \\
					Sensitivity       & 0.7976 & 0.7308 &  & 0.7459 & 0.6777 \\
					Specificity       & 0.7091 & 0.7142 &  & 0.6255 & 0.6464 \\ \hline
				\end{tabular}
				\begin{tablenotes}[flushleft]
					\item[]Note: ROC AUC: a threshold-independent performance measure ranging from 0.5 (model classify as good as random) to 1 (model perfectly classify). Accuracy: correctly predicted tweets over total tweets. Sensitivity: correctly predicted she-tweets over total she-tweets. Specificity: correctly predicted he-tweets over total he-tweets.
				\end{tablenotes}
			\end{threeparttable}
		}
	\end{centering}	
	\caption{\label{tab:main-results}Model Evaluation on Test Set}
\end{table}

Table \ref{tab:main-results} further illustrates the main result by measures of accuracy, sensitivity and specificity. Accuracy is how many tweets the model correctly classifies.  Sensitivity and specificity investigate how good the model is in either class. Sensitivity is how many she-tweets the model classifies correct over total she-tweets. Specificity is how many he-tweets the model classifies correct over total he-tweets. The unbalanced sample- panel presents that the masked version of the Neural Network model has an accuracy of 77 percent as opposed to the Naive Bayes model that has an accuracy of 75 percent. Also, it shows that the sensitivity of the Neural Network model is 21 percent and the specificity is 96 percent. In other words, of the total she-tweets it classifies 21 percent correctly to the she-class and of the total he-tweets it classifies 96 percent correctly to the he-class. The fact that the masked neural network model mainly use the simple rule of classifying to the he-class is due to having imbalanced classes while deciding to optimize accuracy. There are around three times more he-tweets than she-tweets and by optimizing accuracy the model takes into account that the best guess for any tweet is that it belongs to the he-class. The class imbalance is large in magnitude, out of 100 tweets 74.6 contains \textit{he} and 25.4 contains \textit{she}. The class imbalance does not seem to be specific to Swedish tweets. The \textit{she} -class range from 23 percent on Wikipedia to 43 percent on popular Swedish blogs.\footnote{For examples of Swedish text sources and the class imbalance found in them see appendix table \ref{tab:Class imbalance in other Swedish text sources}. The class imbalance in the Swedish tweets is not due to females being mentioned in a passive sense; inclusion of the words \textit{him} (honom) and \textit{her} (henne) in the two classes only increase the she\textit{-} class by 0.7 percentage points. It is not the case that most twitter users are males that talk about other males as the likelihood of a random Twitter user being male is 56 - 58 percent \citep{ISS2016}.} Similar levels of class imbalance exist in the large corpus of scanned English books available at Google's Ngram Viewer \citep{NgramViewer}.  The class imbalance is not a reflection of gender norms according to my definition of it as a set of soft boundaries defining a female and male identity. I interpret the class imbalance as males being talked about to a larger extent than females are. I choose to optimize accuracy because I want to capture the gender norms through a continuous gendered language- variable while taking into account that the bulk of tweets is about men. The class imbalance is by itself a result which should not be understated. Later on in the paper, the masked version of the Neural Network model, for which accuracy is optimized, evaluates the Metoo event.

The choice of optimizing accuracy withstanding, it is not a convenient
standpoint for illustrating the performance of the model. Table \ref{tab:main-results}, the Balanced Sample- panel, illustrates that the model captures gendered language, and not simply predicts all tweets to belong to the he-class. The Balanced sample- panel shows results on a balanced sample for the choice of maximizing the predictive power of the model in either class. The balanced sample is created by randomly throwing away excess he-tweets from the test set until an equal count of he- and she-tweets exists. The model is trained with a binary variable representing \textit{he} and \textit{she}, but I consider gendered language be a continuous variable ranging from 0 (male) to 1 (female). The training against the he/she-variable is a method to capture the underlying unobserved continuous gendered-language variable. The output of the model is a probability distribution since the last stage of the model is specified as a logistic function (just as for a Logit model). To classify into either class, a specific threshold value is specified at the probability distribution. The ROC AUC-score is the preferred performance metric since it is independent of any threshold chosen. In the preceding section was the threshold chosen to optimize accuracy, but the threshold can be chosen in many different ways. The balance sample- panel presents results from a threshold optimized for making the model being equally good at predicting to the two class. Table \ref{tab:main-results} presents that the masked version of the Neural Network model now has an accuracy of 68 percent. At the same time, sensitivity is at 75 percent and specificity at 62 percent. In other words, when making a fictive reality where tweets are as much about women as men, the model predicts she-tweets correctly 75 percent of the time.

Table \ref{tab:highPredictingWords} illustrates which words that
the masked Neural Network model uses to predict.  The model use sentences as input and not words in isolation, which is hard to visualize in a table. To easier illustrate what the model does the median predicted probability ranks the words, termed Word Color (WC) by me since each word ``get colored'' by the prediction for the tweets to which it belongs. The higher the WC the more often is the word used in a tweet classified to the she-class. The method is conceptually equivalent to evaluating a Logit model on the sample mean, except for that I strive to preserve the ordering of words taken into account by the model by using the median of the predicted probability.\footnote{Another way of understanding what a neural network model does is to
	train a generative network or include an attention mechanism in the
	network. However, a generative network will still yeild choosen examples
	of what it found, for example the ``mean'' sentence of a tweet containing
	the word \textit{he}. An attention mechanism would only illustrate
	how the network predicts at individual tweets. Thus, at present I
	refrain from making such extentions.} The table displays the
most frequent words over the predicted probability distribution.  For example, the masked word \textit{grandfathergandmother (morfarmormor)} has a word color of 0.77, meaning that the word is found in tweets predicted to belong to the she-class. Table   \ref{tab:highPredictingWords} displays that words related to sport are found in tweets predicted to be about a male. Words related to personal relations are found in tweets predicted to be about a female, even though the words are masked.\footnote{The outmost left tail in table \ref{tab:highPredictingWords}, where
	we find \textit{credit}, \textit{free} and \textit{casino}, is due
	to very few words having a median predicted probability (WC) close
	to one. The table also display the word \textit{the lady} (\textit{tanten}).
	One could argue that this word should have been masked just as \textit{the
		ladies} (\textit{tanter}), \textit{lady} (\textit{tant}) and \textit{neighborhood
		lady} (\textit{granntanter}) are (seen in the table of words used for masking in the online appendix). That only one word on the list in table \ref{tab:highPredictingWords}
	denotes a gendered subject shows that most relevant words indeed are
	masked.} \footnote{One might want to control which words that are relatively more used in either class by evaluating on the balanced sample, as opposed to absolutely used in the unbalanced sample. Appendix table \ref{tab:highPredictingWords-balanced} replicates the same exercise on a balanced sample and shows that the pattern of sports versus personal relations still holds.}

\begin{small}
	\begin{center}
		\begin{longtable}{lll} 
			\caption{\label{tab:highPredictingWords}Frequent words over the predicted probability distribution for the masked model} \\
			&           &         \\ 
			\hline \hline
			WC   & English   & Swedish  \\ 
			\hline
			\endfirsthead
			\multicolumn{3}{c}%
			{\tablename\ \thetable\ -- \textit{Continued from previous page}} \\
			\hline \hline
			WC   & English                                         & Swedish                                  \\ \hline
			\endhead
			\hline \hline \multicolumn{3}{r}{\textit{Continued on next page}} \\
			\endfoot
			\hline \hline
			\multicolumn{3}{l}{Note: The most “male” word with a WC of close to zero is} \\
			\multicolumn{3}{l}{at the top of the table and the most “female” word with a} \\
			\multicolumn{3}{l}{WC close to one is at the bottom of the table. The table is} \\
			\multicolumn{3}{l}{generated in the following way: The median predicted} \\
			\multicolumn{3}{l}{probability for each word is calculated from the predicted} \\
			\multicolumn{3}{l}{probability of each tweet, named Word Color (WC). The WC of} \\
			\multicolumn{3}{l}{all words is binned into 20 groups and for each quantile is} \\
			\multicolumn{3}{l}{the 5 most frequent words displayed. The table is generated} \\
			\multicolumn{3}{l}{on the test set and evaluated by the masked neural network} \\
			\multicolumn{3}{l}{model. The translation from Swedish is made by the author.} \\
			\endlastfoot
			0.01 & contract                                        & kontrakt                                 \\
			0.02 & united                                          & united                                   \\
			0.03 & games                                           & matcher                                  \\
			0.03 & player                                          & spelare                                  \\
			0.04 & club                                            & klubb                                    \\
			0.05 & the season                                      & säsongen                                 \\
			0.06 & goal                                            & mål                                      \\
			0.07 & the game                                        & matchen                                  \\
			0.08 & game                                            & match                                    \\
			0.09 & play                                            & spela                                    \\
			0.10 & the boll                                        & bollen                                   \\
			0.11 & team                                            & lag                                      \\
			0.13 & played                                          & spelade                                  \\
			0.14 & football                                        & fotboll                                  \\
			0.15 & penalty                                         & straff                                   \\
			0.16 & field                                           & plan                                     \\
			0.18 & ready                                           & klar                                     \\
			0.18 & play                                            & spelar                                   \\
			0.18 & miss                                            & missar                                   \\
			0.20 & em                                              & em                                       \\
			0.21 & europe                                          & europa                                   \\
			0.23 & the chance                                      & chansen                                  \\
			0.23 & worse                                           & sämre                                    \\
			0.25 & president                                       & president                                \\
			0.25 & \textless{}familyname\textgreater{}             & \textless{}familyname\textgreater{}      \\
			0.27 & short                                           & kort                                     \\
			0.27 & leave                                           & lämnar                                   \\
			0.27 & last                                            & förra                                    \\
			0.28 & score                                           & poäng                                    \\
			0.30 & bad                                             & dålig                                    \\
			0.32 & smallest                                        & minst                                    \\
			0.32 & latest                                          & senaste                                  \\
			0.34 & manage                                          & lyckas                                   \\
			0.34 & still there                                     & kvar                                     \\
			0.35 & before                                          & före                                     \\
			0.35 & against                                         & mot                                      \\
			0.37 & were                                            & varit                                    \\
			0.39 & good                                            & bra                                      \\
			0.40 & fuck                                            & fan                                      \\
			0.40 & in                                              & in                                       \\
			0.43 & like                                            & ju                                       \\
			0.43 & come                                            & kommer                                   \\
			0.44 & in                                              & i                                        \\
			0.44 & had                                             & hade                                     \\
			0.44 & .                                               & .                                        \\
			0.45 & \textless{}user\textgreater{}                   & \textless{}user\textgreater{}            \\
			0.46 & ,                                               & ,                                        \\
			0.47 & \_\_\_                                          & \_\_\_                                   \\
			0.48 & are                                             & är                                       \\
			0.48 & at                                              & att                                      \\
			0.50 & \textless{}hashtag\textgreater{}                & \textless{}hashtag\textgreater{}         \\
			0.51 & and                                             & och                                      \\
			0.52 &                                                 &                                          \\
			0.53 & I                                               & jag                                      \\
			0.54 & \textless{}url\textgreater{}                    & \textless{}url\textgreater{}             \\
			0.55 & you                                             & dig                                      \\
			0.57 & love                                            & älskar                                   \\
			0.58 & me                                              & mig                                      \\
			0.59 & my                                              & mitt                                     \\
			0.59 & \textless{}boygirl\textgreater{}                & \textless{}killetjejkillle\textgreater{} \\
			0.61 & my                                              & mina                                     \\
			0.62 & *heart eyes*                                    & *heart eyes*                             \\
			0.63 & children                                        & barn                                     \\
			0.63 &                                                 &                                          \\
			0.64 & *heart*                                         & *heart*                                  \\
			0.66 & \textless{}womenmen\textgreater{}               & \textless{}kvinnormän\textgreater{}      \\
			0.66 & \textless{}thedaughtertheson\textgreater{}      & \textless{}dotternsonen\textgreater{}    \\
			0.66 & my                                              & min                                      \\
			0.67 & \textless{}brothersister\textgreater{}          & \textless{}brorsyster\textgreater{}      \\
			0.69 & \textless{}sondaugther\textgreater{}            & \textless{}dotterson\textgreater{}       \\
			0.71 & friend                                          & \textless{}väninnavän\textgreater{}      \\
			0.72 & \textless{}mum'sdad's\textgreater{}             & \textless{}mammaspappas\textgreater{}    \\
			0.72 & \textless{}grandfathergrandmother\textgreater{} & \textless{}farfarfarmor\textgreater{}    \\
			0.74 & beautiful                                       & vacker                                   \\
			0.75 & \textless{}mumdad\textgreater{}                 & \textless{}mammapappa\textgreater{}      \\
			0.76 & fi                                              & fi                                       \\
			0.77 & \textless{}thewomenthemen\textgreater{}         & \textless{}kvinnornamännen\textgreater{} \\
			0.77 & \textless{}grandfathergrandmother\textgreater{} & \textless{}morfarmormor\textgreater{}    \\
			0.78 & pink                                            & rosa                                     \\
			0.80 & yearly                                          & åriga                                    \\
			0.81 & makeup                                          & smink                                    \\
			0.81 & nails                                           & naglar                                   \\
			0.82 & hagen                                           & hagen                                    \\
			0.82 & the cookie                                      & kakan                                    \\
			0.82 & pippi                                           & pippi                                    \\
			0.86 & noora                                           & noora                                    \\
			0.87 & book                                            & book                                     \\
			0.87 & thatcher                                        & thatcher                                 \\
			0.88 & raped                                           & våldtagen                                \\
			0.90 & zara                                            & zara                                     \\
			0.90 & dress                                           & klänning                                 \\
			0.91 & pregnant                                        & gravid                                   \\
			0.92 & veil                                            & slöja                                    \\
			0.94 & the lady                                        & tanten                                   \\
			0.94 & \textless{}hishers\textgreater{}                & \textless{}hanshennes\textgreater{}      \\
			1.00 & lift                                            & lyft                                     \\
			1.00 & credit                                          & credit                                   \\
			1.00 & free                                            & free                                     \\
			1.00 & casino                                          & casino                                   \\
			1.00 & code                                            & code                                    \\
		\end{longtable}
	\end{center}
\end{small}

\section{Did \#metoo change gender norms in Sweden?}

\subsection{The \#Metoo movement in Sweden}

The Metoo movement started in the US and quickly spread to Sweden. In the morning of the 17th October 2017, people started to share and read the metoo-post on social media: 

\begin{quote}
	\#MeToo. 
	
	If all the women who have been sexually harassed or assaulted wrote "Me too." as a status, we might give people a sense of the magnitude of the problem. 
\end{quote}

\begin{figure}[H]
	\begin{centering}
		\includegraphics[width=1\textwidth]{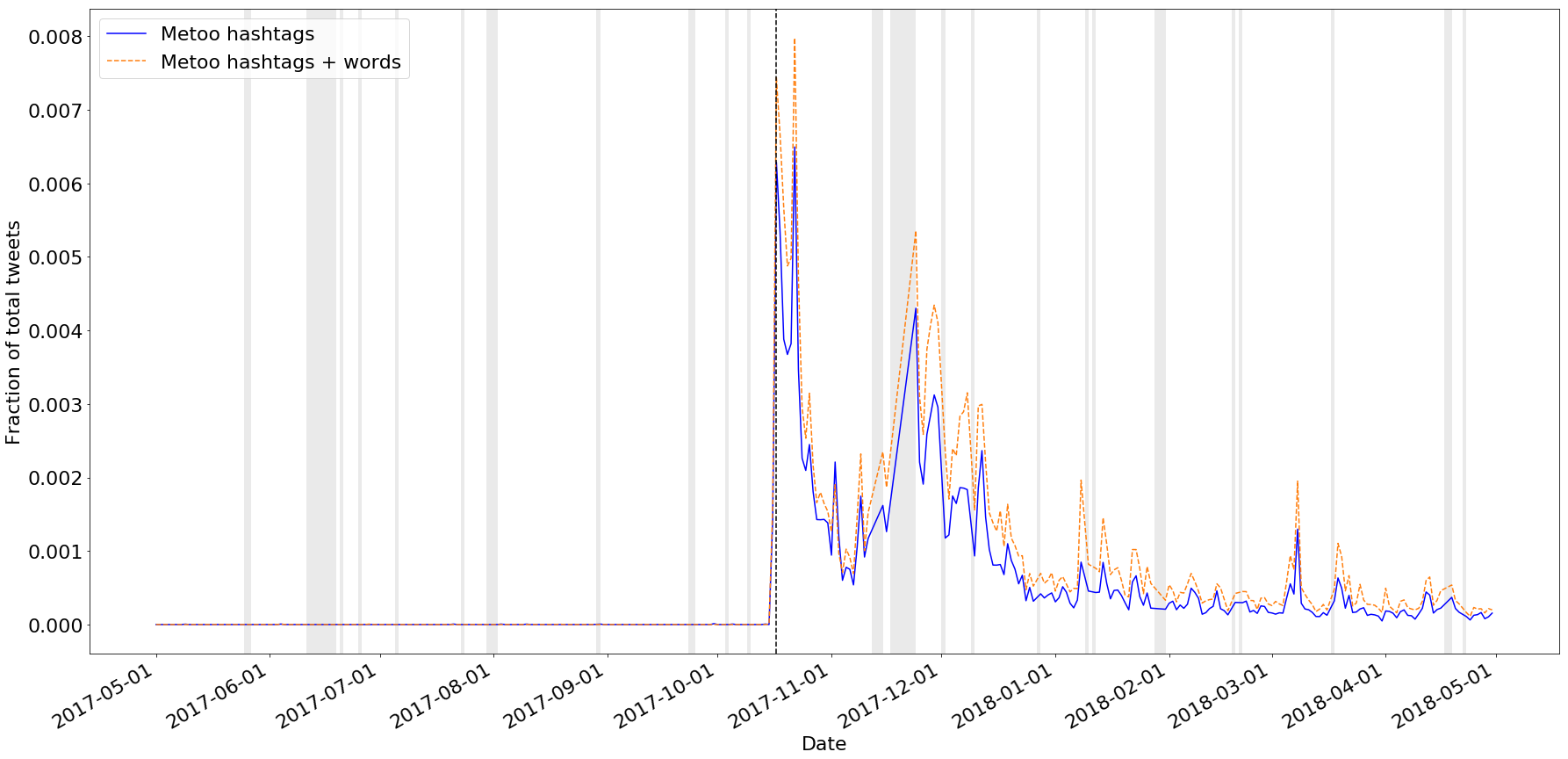}
		\par\end{centering}
	\caption{\label{fig:The-metoo-shock}The \#Metoo event}
	\begin{flushleft}
		\footnotesize{Note: The ``metoo hashtag''-series includes appearances of the hashtag. The ``metoo hashtag + word''- series in addition includes appearances of the word \textit{metoo}, not prefixed with \#. The shaded regions represent missing dates.}
	\end{flushleft}
\end{figure}

During the following hours, people shared more and more metoo-posts. Figure \ref{fig:The-metoo-shock} displays the Metoo movement on Swedish Twitter. The 17th of October 2017, 0.63 percent of all Swedish non-retweeted tweets contains the hashtag.\footnote{I choose to present the Metoo movement with total tweets in the denominator as the event in itself is correlated with a rise in the share of hashtagged tweets, which might be due to the event in itself. Those considerations withstanding, the previously quoted figure for the 17th of October is equivalent to 4 percent of all hashtagged tweets containing the metoo-hashtag.} The hashtag peaks the 22th of October at 0.65 percent of all tweets. As displayed in the figure there is virtually no appearances of the hashtag before the 16th of October 2017. Specifically, there is a total of 10 appearances of the hashtag for the five months before the Metoo event. This paper argues that the Metoo event's exact date of occurrence, the 17:th of October, is exogenous in regards to the following of gender norms. The Metoo movement continued the subsequent weeks after the initial event in Sweden with women in different sectors of the Swedish economy collecting their experiences at the workplace under different hashtags.\footnote{To give examples, \#teknisktfel, \#tystnadtagning and \#medvilkenrätt (\#technicalerror, \#silencetake, \#bywhichright).} Figure \ref{fig:The-metoo-shock} also shows a leveling out. For the last included month, April 2018, on average 0.028 percent of all tweets contain the hashtag.

The Metoo event and the following movement raised general awareness of the sexualization of women. This might lead to people changing their gender norms in regards to women and, thus, how they express themselves in text. The discussions following the event spilled over to gender issues in general. Such debates might lead to persons wishing to categorize people as men and women to a lesser extent. The persons participating in the movement most likely hoped for a less gender-stereotypical environment, but it is possible that it became more gender-stereotypical as well.

There are likely differences in how persons of different ages experienced the Metoo event. The younger generation is more prone to consume social media. Younger persons are more likely impacted by having friends and acquaintances sharing experiences of sexual harassment online.
In comparison, the older generation to a greater extent consumes traditional media, such as TV news broadcasts and newspapers.  Without a presence on social media, the impact of the Metoo event is more likely the notion of media personalities being accused of sexual harassment.
The younger share of the population is more likely to have a strong experience of the Metoo event.

This paper captures the norms and the effect on Swedish twitter users. According to a survey performed by \citet{ISS2016} 18 percent of Internet users use Twitter and 8 percent post tweets in Sweden. The age distribution is skewed: the highest share of Twitter users is among 16 to 25-year-olds at 35 percent and declines with age. Men use Twitter more than females, the likelihood of a random Twitter user being male is 56 to 58 percent. Hence, the gender norms the younger generation holds are defined and investigated, with a slight bias towards men.

\subsection{\label{subsec:Research-design}Research design}

The main dependent variable \textit{Follow Norms} evaluates if gender norms changes in relation to the Metoo movement. The exact model version used for the evaluation is the masked version of the neural network model where accuracy has been optimized, henceforth called Gini for short. \textit{Follow Norms} is a binary indicator on if the Geni predicts a tweet $i$ correct:

\begin{equation} \label{eq:followNormsdef} 
Follow Norms_{i}= \begin{cases}
1 & \text{if } Prediction_i=TrueGender_i \\
0 & \text{if } Prediction_i\neq TrueGender_i 
\end{cases}
\end{equation}

An average of \textit{Follow Norms} for a sample of tweets is equivalent to the predictive accuracy of Geni. The variable \textit{Follow Norms} indicates if the norms in Year 1 are followed because Geni is trained on data from this year. In other words, norm changes are benchmarked against a definition of the gender norms obtained in Year 1.  

\begin{figure}[H]
	\begin{centering}
		\includegraphics[width=0.8\textwidth]{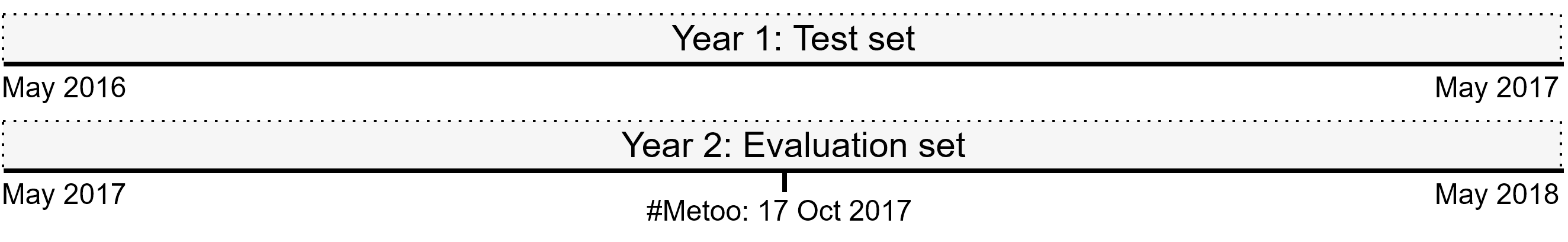}
		\par\end{centering}
	\centering{}\caption{\label{fig:Research-design}Research design}
\end{figure}

The Metoo event occurs on the 17th of October 2017. To be sure not to capture the effect of the Metoo movement leading to a more intensive debate about gender tweets containing the metoo-hashtag and other correlated hashtags are removed.\footnote{This cuts the sample size to 97-98 percent of the original.} Although the event's exact date of occurrence is exogenous, identification relies on having a counterfactual scenario. Figure \ref{fig:Research-design} illustrates the baseline research design. There might exist time trends in norm following. The tweets of Year 2 (where the Metoo event takes place) allow us to investigate possible time trends five months before the event. There might also exist seasonal variation in the following of gender norms. We can use the tweets of year 1 (the previous year) as a comparison group to approximate such seasonal variation by calendar day fixed effects. Only the test set of Year 1 functions as a comparison group because Geni is not trained on those tweets. They better approximate the predictive accuracy of Geni, the level of \textit{Follow Norms}, which one expects in Year 2 in the absence of any effect. 

The baseline specification becomes:

\begin{equation} \label{eq:baseline} 
Follow Norms_{itu}=Year2_{t}+ After Metoo_{t}+v_t+u_{itu}
\end{equation}

were each observation is a tweet $i$ posted at calender day $t$ by user $u$. The variable $Year2$ is a binary indicator on if the tweet belong to the Year 2. The main independent variable \textit{After Metoo} is a binary indicator on if the tweet is posted in Year 2 after the Metoo event. The tweets of Year 2 are compared to the tweets from the test set of Year 1 with calendar days time fixed effects $v_{t}$ to control for seasonal variation. Time is indexed to start at 2017-05-01 for both years, and thus each calendar day has its own intercept (for example, 2016-05-01 and 2017-05-01 has one intercept). 

The set-up used in this paper is a difference-in-difference setup, but compared to many other cases when this research design is used many timepoints exist (large t). It is also similar to regression discontinuity in time (RDIT). As norms are supposed to be slowly changing, the approximately 5-6 month before and after the Metoo event can be seen as the window size used for the RDIT-estimate. There is an ongoing discussion on how to cluster standard errors for similar settings as the current one \citep{bertrand2004much,abadie2017when}. The approach taken in this paper is to present conservative estimates, although they might be unnecessarily harsh. For specifications including calender day fixed effects or time trends standard errors are clustered on calender days. I choose to present the standard errors clustered per calender day in the baseline specification since it provides a framework which easier extend to user fixed effects. 

In robustness checks, autocorrelation in the residuals are taken into account as one could argue for calendar day- clustered standard errors being problematic since they miss such variation. The data is aggregated per calender day, $\bar{Y}_{tg}=\sum_{i \in group\ g}Y_{itug}/I$ where $g$ refers to group (Year 1 or 2). With the transformed data, the baseline specification (\ref{eq:baseline}) becomes $(\bar{Y}_{t,g=Year2}-\bar{Y}_{t,g=Year1})=\ c+AfterMetoo_{t}+(u_{t,g=Year2}-u_{t,g=Year1})$ and autocorrelation is estimated by Newey-West standard errors with a lag length of 4 (a bandwidth of 5). Since the data is aggregated weights, equal to the sum of observations for the two groups, are applied during estimation. 

In additional robustness specifications, I extend the baseline specification (\ref{eq:baseline}) with user fixed effects. This is feasible since each tweet contains a user id. Standard errors are clustered on calender days and users with the correction to the variance-covariance matrix estimator presented in \citet{cameron2011robust}. I choose the baseline specification without including user fixed effects since I want to investigate the change in tweets. If users stop posting tweets due to the event third-party persons are less affected by the norms no longer followed, and I would like to capture such composition effects. There is no \textit{a priori} reason to expect users to leave Twitter for another online forum, but the user fixed effects controls for alike possibilities.

In the robustness analysis a second control strategy is also employed. Theoretically, Geni should perform worse over time only since users introduce new words and phrases for describing the same thing as the words and phrases she already knows. At first-hand this is not a great worry because Geni's accuracy on average increases in Year 2 as compared to the test set of Year 1 (not shown). Another model has been trained to predict the words \textit{I} and \textit{we} in the equivalent way that Geni was trained to predict \textit{he} and \textit{she}.\footnote{See appendix A for details.} The I/We-series approximates a togetherness-norm, it captures words and phrases depicting what people do alone or together. The Togetherness-series function as a placebo test, it is not expected to change due to the Metoo movement. The series mainly control for the evolution of the  Swedish language. In the analysis, a decrease in Geni's accuracy, the \textit{Follow Norms}- variable, is interpreted as norms being less followed but it might just arise since the language evolve. The Togetherness-series will respond in the same way and is therefore used to filter out possible language evolution- effects.

\subsection{Results}

\begin{table}[H]
	\begin{centering}
		\noindent\resizebox{0.7\textwidth}{!}{%
			\begin{threeparttable}
				\begin{tabular}{lcccc}
					\hline \hline
					& Raw          & Baseline     & Cutoff independence & Female Focus \\
					& (1)          & (2)          & (3)                 & (4)          \\
					Dep Var     & Follow Norms & Follow Norms & Gendered Language   & She Count    \\ \hline
					&              &              &                     &              \\
					After Metoo & -0.003***    & -0.015***    & 0.013***            & 0.022***     \\
					& (0.001)      & (0.005)      & (0.004)             & (0.007)      \\
					&              &              &                     &              \\
					&              &              &                     &              \\
					Time FE     & No           & Yes          & Yes                 & Yes          \\
					User FE     & No           & No           & No                  & No           \\
					SE:s        & Robust   & Clustered    & Clustered           & Clustered    \\
					&              & on day       & on day              & on day       \\
					&              &              &                     &              \\
					Tweets      & 989 028      & 1 192 719    & 1 192 719           & 1 192 719    \\
					Days (T)    & 319          & 365          & 365                 & 365          \\
					Users (N)   & 95 025       & 114 040      & 114 040             & 114 040      \\
					&              &              &                     &              \\
					Data        & Year2        & Year2/1      & Year2/1             & Year2/1    \\ \hline \hline
				\end{tabular}			
				\begin{tablenotes}[flushleft]
					\item[]Note: {*}{*}{*} p$<$0.01, {*}{*} p$<$0.05, {*} p$<$0.1.
				\end{tablenotes}
			\end{threeparttable}
		}
	\end{centering}	
	\caption{\label{tab:metoo-without-user-fe}Metoo results without user fixed
		effects}
\end{table}

Table \ref{tab:metoo-without-user-fe} presents the main result on weather the Metoo event changes gender norms on Swedish Twitter. The first column presents a raw estimate of -0.003 showing that gender norms on average are less followed in tweets six months after the Metoo event compared to five months before. The magnitude is small, attributing the decrease of 0.3 percentage points in \textit{Follow Norms} to the Metoo movement, the estimate indicates the following of gender norms decreasing by 0.4 percent over the sample mean of the five months before. The implicit counterfactual in the comparison is the average level of \textit{Follow Norms} in the five months before.  Although the  \textit{Follow Norms}- series does not exhibit pre-trends (appendix table \ref{tab:metoo-with-time-trends}), the estimate might be biased. There might be seasonal variation in gender norms adherence, which would have changed norm following even in the absence of the Metoo movement.

The second column of table \ref{tab:metoo-without-user-fe} displays the baseline specification. The test set of Year 1 function as a comparison group. Seasonal variation is filtered out by allowing the same calendar day to have its own effect and only variation around the calendar day means contribute to the estimate of \textit{After Metoo}. The coefficient of -0.015 shows gender norms on average being less reflected in tweets after the event. The magnitude of the estimate is fairly large, 1.9 percent of the sample mean before the event. Extrapolationg the difference implies that tweets would not reflect any of the gender norms followed before metoo after 18 Metoo movements (based on counting a level of \textit{Follow Norms} of 50 percent as indicating no adherence to past gender norms, since this is the level achieved when Geni has lost all predictive power). Geni does not make any discrepancy between wanted and unwanted gender norms, and the comparison is not normative.

Like any dependent variable \textit{Follow Norms} include measurement errors. For the present case, additional errors are introduced in the \textit{Follow Norms}- variable because Geni binarizes the probability for a tweet to be about a male or a female (a continuous variable from 0 to 1), by having a specific cutoff for predicting to either gender. The specific cutoff yields the predicted value, needed to create \textit{Follow Norms}. To verify the main result, table \ref{tab:metoo-without-user-fe} column (3) presents estimates with the probability as a dependent variable, which includes fewer measurement errors. The variable does not carry the same interpretation as Follow Norms: it measures the degree of genderedness in a tweet and I therefore call it \textit{Gendered Language}. The estimate of 0.013 shows that the language in the tweets after the Metoo event is congruent with it being about a female to a higher degree. The magnitude is fairly large, 3.0 percent of the sample mean. If Geni went 45 years into a future, she would think that all tweets are about women without any hesitation. A direction towards “more female” gendered language does not mean that the tweets are about females, it can equally well describe men. 

The \textit{Follow Norms}- variable implements a cutoff making Geni good at predicting independent of the choice of writing either he or she since it is not included in the operationalized definition of the gender norms. If there were many more appearances of she-tweets after metoo the model would still predict those tweets to belong to the she-class, conditional on there being the same level of gendered language in the additional she-tweets, and the \textit{Follow Norms}- variable would indicate adherence to gender norms. One could argue that it would be beneficial for the model to take into consideration how much tweets mentions males and females. Instead of including such aspects in the definition of gender norms, I choose to present results on the count of she-tweets as a way to measure which gender the tweet is about. Table \ref{tab:metoo-without-user-fe}, column (4) shows that there are 2.2 percentage points more she-tweets after the Metoo event than before. This translates to an effect size of 9.2 percent of the sample mean. Before metoo 76.3 percent of all tweets are he-tweets, and the estimate indicates that it would take 35 Metoo movement to replace all he-tweets with she-tweets on Twitter. The large volume of talk about men is decreasing. 

\begin{table}[H]
	\begin{centering}
		\noindent\resizebox{0.7\textwidth}{!}{%
			\begin{threeparttable}
				\begin{tabular}{lcccc}
					\hline \hline
					& She               & He                & She          & He           \\
					& (1)               & (2)               & (3)          & (4)          \\
					Dep Var     & Gendered Language & Gendered Language & Follow Norms & Follow Norms \\ \hline
					&                   &                   &              &              \\
					After Metoo & 0.013***          & 0.007*            & 0.020***     & -0.005***    \\
					& (0.003)           & (0.004)           & (0.005)      & (0.001)      \\
					&                   &                   &              &              \\
					Time FE     & Yes               & Yes               & Yes          & Yes          \\
					User FE     & No                & No                & No           & No           \\
					SE:s        & Clustered         & Clustered         & Clustered    & Clustered    \\
					& on day            & on day            & on day       & on day       \\ 
					&                   &                   &              &              \\
					Tweets      & 293 869           & 898 850           & 293 869      & 898 850      \\
					Days (T)    & 365               & 365               & 365          & 365          \\
					&                   &                   &              &              \\
					Data        & Year2/1           & Year2/1           & Year2/1      & Year2/1      \\ \hline \hline
				\end{tabular}
				\begin{tablenotes}[flushleft]
					\item[]Note: {*}{*}{*} p$<$0.01, {*}{*} p$<$0.05, {*} p$<$0.1. 
				\end{tablenotes}
			\end{threeparttable}
		}
	\end{centering}	
	\caption{\label{tab:metoo-subset-he-she}Metoo results estimated separately
		for tweets about females and males}
\end{table}

Table \ref{tab:metoo-subset-he-she} shows the baseline specification for the subsets of she- and he-tweets separately. Column (1) and (2) show that for both she- and he-tweets gendered language increases: both women and men are more described as women were the previous year. Column (3) shows that for she-tweets this, on average, entails following the norms found in Year 1 to a greater extent. Column (4) presents that for the he-tweets the increase in gendered language, on average, is congruent with following the norms found in Year 1 to a smaller extent. I hypothesized that the Metoo movement lead to a norm shift away from stereotyping the genders, implying that women should be described more as men were the previous year, with a resulting decrease in the following of past gender norms for she-tweets. Instead, the overall gender norm shifts towards describing both genders more as females were the previous year. After some thought, the finding is quite reasonable, \textit{a priori} there is no reason why a norm shift implies convergence between the genders. The finding is possible because the operationalized definition of gender norms does not include which gender the tweet, in reality, is about. The current table points to males being described less according to past gender stereotypes as contributing to the baseline effect. The estimate of -0.005 only indicates an effect size of 0.5 percent of the sample mean before the event.  Even though he-tweets contributes more to the average baseline effect since they around three times more, the size of the baseline effect of -0.015 in table \ref{tab:metoo-without-user-fe}, column (2), cannot only be explained by he-tweets being less gender stereotypical. The current table also points to she-tweets driving the baseline effect, but in the "wrong" direction. The baseline effect might arise since there are more she-tweets for which norms are less followed due to the event, but females on average are pictured more according to past gender norms after metoo.

Compared to previous authors I measure gender norms in a much more comprehensive way and it might be that some specific dimension of the gender norms has changed, for example, the tweets might sexualize women less. To investigate if any such patterns exist Table \ref{tab:highestDiff} presents the words which experienced the largest change in accuracy, i.e. in \textit{Follow Norms}. The top of the table presents the top of the differences; Geni predicts those specific words more correctly in Year 1 and more incorrectly after the Metoo event in Year 2. The method employed is for illustrative purposes only, no formal tests are conducted on the words.  A word is dependent on other words surrounding it, and thus, the main method employed in this paper is to model tweets and conduct inference on them.  In table \ref{tab:highestDiff}, the “Diff”-columns displays the change in accuracy by calculating the mean predictive accuracy for each word. Due to this reduction, and since Geni might not use a specific word to a high degree, the table is noisy. I can infer what the model uses to predict on and analyze the change by analyzing the underlying tweets. \footnote{A longer version of the table is available in the online appendix (in Swedish). As far as I understand, I cannot redistribute the tweets according to the terms and services for the Twitter API.} For example, the word \textit{cutie} (sötis) marks a gender stereotype insofar that males are called cutie (with a slightly derogatory tone) when they do specific things, and in Year 2 females are called \textit{cutie} as well. The *Beating Heart- emoji* is used in combination with other words, such as being nice or able, by males and females mostly for describing their partners. In Year 2, the pattern changes: the *Beating Heart- emoji* is used in combination with qualities that the model do not predict correct, such as males being \textit{very good} (jätteduktig). Relating to the word \textit{views} (åsikterna), \textit{to have views} (ha åsikter) is encoded as indicating a male gender stereotype, it is mostly males that have views in Year 1, but in Year 2 women are described as having views to a higher degree.\footnote{The words \textit{highly pregnant} (höggravid) and \textit{pregnant} (gravid) unsurprisingly describes females to a high degree. In year 2, the two words are used as an explanation for why a woman experiences an action performed by a man, a pattern the model has not learned. Insofar that pregnant indicate a gender stereotype by just being a state of the world without having any explanatory power, the pattern still indicate a change.} The above are selected examples, but indicates that there not are one clear-cut dimension of the gender norms that changes.

\begin{small}
	\begin{center}
		\begin{longtable}{lll}
			\caption{\label{tab:highestDiff}Change in accuracy on word- level} \\
			&                                        &                                \\ \hline \hline
			Diff  & English                                        & Swedish                                   \\ \hline
			\endfirsthead
			\multicolumn{3}{c}%
			{\tablename\ \thetable\ -- \textit{Continued from previous page}} \\
			\hline \hline
			Diff  & English                                        & Swedish                                   \\ \hline
			\endhead
			\hline \hline \multicolumn{3}{r}{\textit{Continued on next page}} \\
			\endfoot
			\hline \hline
			\multicolumn{3}{l}{Note: The top of the table shows words included in tweets that were}      \\
			\multicolumn{3}{l}{correctly predicted before the metoo-event and became incorrectly}        \\
			\multicolumn{3}{l}{predicted afterward. The table is generated in the following way; The}    \\
			\multicolumn{3}{l}{data included is five months of the test set, from 2016-10-18 and}        \\
			\multicolumn{3}{l}{forward, and five months from the evaluation set, from 2017-10-18}        \\
			\multicolumn{3}{l}{and forward. For each of the periods the mean accuracy of each word is}   \\
			\multicolumn{3}{l}{calculated. The difference, named “Diff” in the table, is formed by}      \\
			\multicolumn{3}{l}{subtracting the evaluation set period's mean accuracy from the test}      \\
			\multicolumn{3}{l}{set's mean accuracy. The difference is discretized to 20 groups of 0.005} \\
			\multicolumn{3}{l}{steps- intervals and the 10 most frequent words from the test set are}    \\
			\multicolumn{3}{l}{selected for each group, as long as the word is included in at least}     \\
			\multicolumn{3}{l}{5 tweets for both periods. Due to space limitations, the table displays}  \\
			\multicolumn{3}{l}{only the top of the difference distribution. The full table and}          \\
			\multicolumn{3}{l}{underlying tweets in Swedish are included in the online appendix. The}    \\
			\multicolumn{3}{l}{translation from Swedish is made by the author.}                         \\
			\endlastfoot
			0.78 & highly pregnant                                & höggravid                                \\
			0.73 & inspiration source                             & inspirationskälla                        \\
			0.71 & coco                                           & coco                                     \\
			0.68 & prisoner                                       & fånge                                    \\
			0.63 & cutie                                          & sötis                                    \\
			0.63 & main                                           & main                                     \\
			0.63 & pregnant                                       & gravida                                  \\
			0.63 & handelbars                                     & styret                                   \\
			0.61 & N                                              & N                                       \\
			0.60 & the univers'                                   & universums                               \\
			0.60 & nuv                                            & nuv                                      \\
			0.58 & scared to death                                & skräckslagen                             \\
			0.57 & paranoid                                       & paranoid                                 \\
			0.57 & =(                                             & =(                                       \\
			0.57 & put forward                                    & framställa                               \\
			0.55 & cute                                           & \textless{}gulligegulliga\textgreater{}  \\
			0.55 & fringe                                         & lugg                                     \\
			0.55 & thai                                           & thai                                     \\
			0.54 & remix                                          & remix                                    \\
			0.53 & k-g                                            & k-g                                      \\
			0.53 & horse                                          & horse                                    \\
			0.52 & jemen                                          & jemen                                    \\
			0.50 & kreml                                          & kreml                                    \\
			0.48 & relate                                         & relatera                                 \\
			0.48 & give joy                                       & glädjer                                  \\
			0.47 & *Beamed Eighth Notes- emoji*                   & *Beamed Eighth Notes- emoji*             \\
			0.46 & quick                                          & hastigt                                  \\
			0.46 & necessary                                      & nödvändig                                \\
			0.45 & draw                                           & tecknar                                  \\
			0.45 & tattooed                                       & tatuerade                                \\
			0.45 & disagreeing                                    & osams                                    \\
			0.45 & missunderstanding                              & missförstånd                             \\
			0.44 & copy                                           & kopierar                                 \\
			0.43 & body guard                                     & livvakt                                  \\
			0.43 & nato                                           & nato                                     \\
			0.43 & *Beating Heart- emoji*                         & *Beating Heart- emoji*                   \\
			0.42 & mean                                           & taskiga                                  \\
			0.42 & svts                                           & svts                                     \\
			0.42 & brainwashed                                    & hjärntvättad                             \\
			0.42 & spokesperson                                   & talesperson                              \\
			0.42 & 06                                             & 06                                       \\
			0.41 & pre-band/dressing of wound                     & förband                                  \\
			0.37 & loaded                                         & laddat                                   \\
			0.37 & foe                                            & ovän                                     \\
			0.37 & views                                          & åsikterna                                \\
			0.37 & cheat                                          & fuskar                                   \\
			0.36 & rinkeby                                        & rinkeby                                  \\
			0.36 & melin                                          & melin                                    \\
			0.36 & babe                                           & babe                                     \\
			0.36 & the dinner                                     & middagen                                 \\
			0.35 & claesson                                       & claesson                                 \\
			0.35 & finance minister                               & finansminister                           \\
			0.34 & \textless{}brother/sister-in-law\textgreater{} & \textless{}svågersvägerska\textgreater{} \\
			0.33 & sexist                                         & sexist                                   \\
			0.31 & wrestling                                      & brottas                                  \\
		\end{longtable}
	\end{center}
\end{small}

\subsection{Robustness checks}

Depending on what one consider the unit of observation to be, one might wish to have different standard errors. So far we have investigated how tweets changes, but one could also wish to consider how the level of norm following per day changes over time. Therefore,  appendix table \ref{tab:metoo-without-user-fe-autocorrelation} and \ref{tab:metoo-subset-he-she-autocorrelation} replicates the preceding two tables \ref{tab:metoo-without-user-fe} and \ref{tab:metoo-subset-he-she}, but displays estimates for daily aggregated data were autocorrelation in the residuals are taken into account by Newey-West standard errors. Now, the baseline estimate is only significant at the 10 percent- level. In general, the standard error increases somewhat, but all except two coefficients retain their significance. One exception is the raw estimate, which only was included in the original table for illustrating the raw data. The other exception is \textit{Gendered Language} for the subset of he-tweets. The p-value for the later is  10.2 percent. The coefficient and the standard errors are equivalent in the two tables, table \ref{tab:metoo-subset-he-she} column (2) and appendix table \ref{tab:metoo-subset-he-she-autocorrelation} column (2). Thus, regardless of the estimation method, the coefficient is close to the 10 percent- significance threshold. In summary, the size of the original standard errors are similar to the size of the Newey-West standard errors.

The selection process behind generating the he/she-sample might be affected by the Metoo event in itself and to address this issue Table \ref{tab:metoo-with-user-fe} replicate the specifications but with user fixed included.\footnote{At first hand, the worry seems unnecessary. I can calculate the day of the first and the last tweet for any user: the entry and the exit rate. On the full sample of tweets, neither the entry nor the exit rate exhibit any immediate apparent trend break around the event, see appendix figure \ref{fig:Entry and exit rates}.} For the Year 2- data, 35 000 users, who have tweeted at least once on either side of the Metoo date, are included, which cut the sample of users to 37 percent of the original. The median included user tweets 8 tweets in Year 2, but some users tweet much more than others rendering the mean much higher. For the baseline specification, the 51 000 users of Year 1 are added regardless of them tweeting at either side of the Metoo event since the tweets of Year 1 are included to allow for time fixed effects. Cutting the sample runs the risk of making it less representative of all users and tweets, but column (1) and (2) replicate the raw and baseline estimate respectively on the cut sample and find them to be very similar to their respective counterparts estimated on the full samples. Column (3) extends the raw estimate by including user fixed effects. Each user is allowed its own intercept and only the variation around a user's mean of \textit{Follow Norms} contribute to the estimate of \textit{After Metoo}. The estimate of -0.005 is lower than the baseline because it shows the average over the post-time period, whereas a U-shaped pattern exists: directly after the initial event users tweet more norm-following tweets, right before Christmas they start to tweet significantly less norm-following and in the middle of March they are back at the pre-event norm-following level (see appendix table \ref{tab:metoo-with-time-trends}, column 3). Column (4) extend the baseline estimate with user fixed effects. The U-shaped pattern disappears when seasonal variation is taken into account (see appendix table \ref{tab:metoo-with-time-trends}, column 4), rendering the average estimate representative. The average individual user decreases his/her level of following norms in the post-period, as compared to how users change their level in Year 1. The estimate of -0.012 is, again, similar to the baseline specification and quite sizable at 1.6 percent of the sample mean before the event. The estimate implies that it would take 22 Metoo movements to change users' tweets not to reflect any gender norms found in Year 1. A comparison of the baseline estimate with and without user fixed effect concludes that most of the average effect steams from users changing their behavior.\footnote{The sample size of the Year 1 tweets is much smaller, 20 percent of the original, because I have subsampled it to allow for model training.  Thus, the calendar day means will be weighted by 80 percent towards the daily means of Year 2, leading to attenuation bias in estimates. In other words, possible trends do not contribute towards the estimate of \textit{After Metoo} to the same degree it would have done if the sample sizes were equal. Weighting the specifications to take into account the inclusion probability of 20 percent for the tweets of Year 1 does not change any of the estimates (not shown).}  

\begin{table}[H]
	\begin{centering}
		\noindent\resizebox{0.7\textwidth}{!}{%
			\begin{threeparttable}
				\begin{tabular}{lcccc}
					\hline \hline
					& Raw: Sample Cut & Baseline: Sample Cut & User FE      & Baseline: User FE \\
					& (1)             & (2)                  & (3)          & (4)               \\
					Dep Var      & Follow Norms    & Follow Norms         & Follow Norms & Follow Norms      \\ \hline
					&                 &                      &              &                   \\
					After Metoo  & -0.003***       & -0.016***            & -0.005***    & -0.012**          \\
					& (0.001)         & (0.005)              & (0.001)      & (0.005)           \\
					&                 &                      &              &                   \\
					Time FE      & No              & Yes                  & No           & Yes               \\
					User FE      & No              & No                   & Yes          & Yes               \\
					SE:s         & Robust          & Clustered            & Clustered    & Clustered on      \\
					&                 & on day               & on user      & day \& user       \\
					&                 &                      &              &                   \\
					Tweets       & 849 518         & 1 053 209            & 849 518      & 1 053 209         \\
					Days (T)     & 319             & 365                  & 319          & 365               \\
					Users (N)    & 35 338          & 65 165               & 35 338       & 65 165            \\
					...in year 2 & 35 338          & 35 338               & 35 338       & 35 338            \\
					...in year 1 &                 & 51 352               &              & 51 352            \\
					&                 &                      &              &                   \\
					Data         & Year2           & Year2/1              & Year2        & Year2/1            \\ \hline \hline
				\end{tabular}
				\begin{tablenotes}[flushleft]
					\item[]Note: {*}{*}{*} p$<$0.01, {*}{*} p$<$0.05, {*} p$<$0.1. 
				\end{tablenotes}
			\end{threeparttable}
		}
	\end{centering}	
	\caption{\label{tab:metoo-with-user-fe}Metoo results with user fixed effects}
\end{table}

Another source of bias can be that Geni performs worse because users introduce new words and phrases that she does not know, which would lead to a decrease in accuracy, in the level of \textit{Follow  Norms}, even in the absence of the Metoo event. So far, we have relied on the test set of Year 1, which Geni was not trained at, for approximating the level. However, the decrease in accuracy due to the language evolving is most likely enhanced by time.  Table \ref{tab:topic-effects} uses the Togetherness-series as a comparison group. We do not expect the main series \textit{Follow  Norms}  and the Togetherness-series to follow the same time trends, and thus such are included in the specifications. Column (1) shows a placebo regression on the Togetherness-series, and displays that the series is stable over time. The result points to no decreases in accuracy due to the language evolving. The rest of table \ref{tab:topic-effects} re-estimate the main specifications because the predictive accuracy can be worse certain days due to the unrecognition of new words and phrases.  Column (2)- (4) present that the previously mentioned U-shaped pattern exists in the main series, as compared to the stable Togetherness-series, after the Metoo event. In column (4) are user fixed effects included in addition to the calendar day fixed effects. A user is counted as two separate users even if the user tweets in both series, since the two different norms might be held separately. The users behind the Togetherness-series are included on the same basis as the main series, i.e. by tweeting on either side of the event. The specification compares how users on average change their tweeting pattern after the event while controlling for decreases in accuracy due to the language evolving within calendar days. Column (4) exhibits the strongest U-shaped pattern. More specifically, the daily averages of how users tweet from the 8 of December to the 28 of March are significantly lower, at the 5 percent- level, than the average level of \textit{Follow Norms} before the event (not shown). The combined result shows that decreases in accuracy due to users introducing new words and phrases do not drive the main results. To use the tweet of Year 1 as a comparison group is preferred since they allow for controlling for seasonal variation and the baseline estimate in table \ref{tab:metoo-without-user-fe} column (2) is the preferred one.

\begin{table}[H]
	\begin{centering}
		\noindent\resizebox{0.7\textwidth}{!}{%
			\begin{threeparttable}
				\begin{tabular}{lcccc}
					\hline \hline
					& Togetherness       & Time Trends        & Time FE            & Time \& User FE    \\
					& (Placebo)          &                    &                    &                    \\
					& (1)                & (2)                & (3)                & (4)                \\
					Dep Var                          & Follow Norms * 100 & Follow Norms * 100 & Follow Norms * 100 & Follow Norms * 100 \\ \hline
					&                    &                    &                    &                    \\
					After Metoo                      & 1.531              & 11.449**           & 10.659*            & 19.153***          \\
					& (1.237)            & (5.646)            & (5.763)            & (5.069)            \\
					After Metoo*t                    & -0.010             & -0.107**           & -0.103**           & -0.150***          \\
					& (0.010)            & (0.048)            & (0.049)            & (0.039)            \\
					After Metoo*t\textasciicircum{}2 & 0.000              & 0.000**            & 0.000*             & 0.000***           \\
					& (0.000)            & (0.000)            & (0.000)            & (0.000)            \\
					&                    &                    &                    &                    \\
					Time FE                          & No                 & No                 & Yes                & Yes                \\
					User FE                          & No                 & No                 & No                 & Yes                \\
					SEs                              & Clustered          & Clustered          & Clustered          & Clustered on       \\
					& on day             & on day             & on day             & day \& user        \\
					&                    &                    &                    &                    \\
					Tweets                           & 1 014 356          & 2 003 384          & 2 003 384          & 1 706 588          \\
					Days (T)                         & 319                & 319                & 319                & 319                \\
					Users (N)                        & 116 376            & 211 401            & 211 401            & 77 863             \\
					&                    &                    &                    &                    \\
					Data                             & Togetherness       & Year2/Togetherness & Year2/Togetherness & Year2/Togetherness \\ \hline \hline
				\end{tabular}
				\begin{tablenotes}[flushleft]
					\item[]Note: The coefficients in
					the table can directly be interpreted as percentage points since the
					binary dependent variable has been multiplied by hundred. The specifications
					interact various indicator variables with time trends. No indicator variables interacted with time is significant more than for \textit{After Metoo}, and hence they are omitted from the table. Specification
					(1): $Y_{it}=1(b_{0}+b_{1}t+b_{2}t^{2})+AfterMetoo(b_{3}+b_{4}t+b_{5}t^{2})+u_{it}$.
					Only for this placebo regression is After Metoo coded to take the
					value one also for the Togetherness-series. Specification (2): $Y_{it}=1(b_{0}+b_{1}t+b_{2}t^{2})+AfterMetoo(b_{3}+b_{4}t+b_{5}t^{2})+HeShe(b_{6}+b_{7}t+b_{8}t^{2})+u_{it}$ where the HeShe-variable is a binary indicator on using the main
					series in the paper. Specification (3): $Y_{it}=AfterMetoo(b_{0}+b_{1}t+b_{2}t^{2})+HeShe(b_{3}+b_{4}t+b_{5}t^{2})+v_{t}+u_{it}$.
					Specification (4): $Y_{it}=AfterMetoo(b_{0}+b_{1}t+b_{2}t^{2})+v_{t}+m_{i}+u_{it}$,
					where a user which tweets in both series is counted as two separate
					users. Aggregating the data to days in specification (1) - (3) and
					estimating Newey-West standard errors with a lag length of 4 does not change the results, except for in specification (2) where the coefficient of the common linear time trend $b_1$ (not shown) becomes significant at the 5- percent level. 
				\end{tablenotes}
			\end{threeparttable}
		}
	\end{centering}	
	\caption{\label{tab:topic-effects}Metoo results comparing to Togetherness}
\end{table}

\section{Conclusion}

This paper interprets the exact date for the Metoo event as exogenous and uses it to evaluate if gender norms can change rapidly. Previous authors present evidence of gender norms being constant over generations \citep{alesina2013origins,fernandez2004mothers,nollenberger2016math}, but this paper finds that gender norms changes in relation to the Metoo movement. More and more conversations take place online which generates text data, and this paper utilizes the newer data source efficiently due to the recent developments in machine learning algorithms. In the paper gender norms are defined in a comprehensive data-driven manner by an LSTM neural network model, allowing for a measure on gender norms experienced by people in their everyday life.

In Swedish tweets six months after the Metoo event, gender norms are on average less adhered to than the five months before. My preferred estimate indicates that 18 Metoo movements would erase adherence to the gender norms that existed the year before. Most tweets are about males, but they start to talk more about females after Metoo. The tweets on average depict both males and females as following female gender roles to a greater extent. The Swedish Twitter sample is not representative of the average gender norm environment in Sweden, and the result is an indication of the possibility for norms to change rapidly.

The quite strong effect found is surprising, and further work should investigate which subpopulations that contribute to it. It would be interesting to investigate if users that tweet more according to gender norms before the Metoo event change more than the once that tweet less according to them. An analysis of which topics contributing tweets cover can also be performed to further the understanding of which type of users that change.

\renewcommand{\bibname}{References}
\addcontentsline{toc}{section}{References}
\bibliographystyle{apa}

\begin{thebibliography}{}
	
	\bibitem[\protect\astroncite{Abadie et~al.}{2017}]{abadie2017when}
	Abadie, A., Athey, S., Imbens, G.~W., and Wooldridge, J. (2017).
	\newblock {When should you adjust standard errors for clustering?}
	\newblock NBER Working Paper 24003, National Bureau of Economic Research.
	
	\bibitem[\protect\astroncite{Akerlof and Kranton}{2000}]{akerlof2000economics}
	Akerlof, G.~A. and Kranton, R.~E. (2000).
	\newblock {Economics and Identity}.
	\newblock {\em The Quarterly Journal of Economics}, 115(3):715--753.
	
	\bibitem[\protect\astroncite{Alesina et~al.}{2013}]{alesina2013origins}
	Alesina, A., Giuliano, P., and Nunn, N. (2013).
	\newblock {On the Origins of Gender Roles: Women and the Plough}.
	\newblock {\em The Quarterly Journal of Economics}, 128(2):469--530.
	
	\bibitem[\protect\astroncite{Arora et~al.}{2016}]{arora2016linear}
	Arora, S., Li, Y., Liang, Y., Ma, T., and Risteski, A. (2016).
	\newblock {Linear Algebraic Structure of Word Senses, with Applications to
		Polysemy}.
	\newblock {\em ArXiv e-prints}.
	
	\bibitem[\protect\astroncite{Bertrand et~al.}{2004}]{bertrand2004much}
	Bertrand, M., Duflo, E., and Mullainathan, S. (2004).
	\newblock {How Much Should We Trust Differences-In-Differences Estimates?}
	\newblock {\em The Quarterly journal of economics}, 119(1):249--275.
	
	\bibitem[\protect\astroncite{Bertrand et~al.}{2015}]{bertrand2015gender}
	Bertrand, M., Kamenica, E., and Pan, J. (2015).
	\newblock { Gender Identity and Relative Income within Households}.
	\newblock {\em The Quarterly Journal of Economics}, 130(2):571--614.
	
	\bibitem[\protect\astroncite{Bisin and Verdier}{2001}]{bisin2001economics}
	Bisin, A. and Verdier, T. (2001).
	\newblock {The Economics of Cultural Transmission and the Dynamics of
		Preferences}.
	\newblock {\em {Journal of Economic Theory}}, 97(2):298--319.
	
	\bibitem[\protect\astroncite{{Bolukbasi} et~al.}{2016}]{BolukbasietAl2016}
	{Bolukbasi}, T., {Chang}, K.-W., {Zou}, J., {Saligrama}, V., and {Kalai}, A.
	(2016).
	\newblock {Man is to Computer Programmer as Woman is to Homemaker? Debiasing
		Word Embeddings}.
	\newblock {\em ArXiv e-prints}.
	
	\bibitem[\protect\astroncite{Bowles et~al.}{2007}]{bowles2007social}
	Bowles, H.~R., Babcock, L., and Lai, L. (2007).
	\newblock {Social incentives for gender differences in the propensity to
		initiate negotiations: Sometimes it does hurt to ask}.
	\newblock {\em Organizational Behavior and Human Decision Processes},
	103(1):84--103.
	
	\bibitem[\protect\astroncite{Buda et~al.}{2018}]{buda2108asystematic}
	Buda, M., Maki, A., and Mazurowski, M.~A. (2018).
	\newblock {A systematic study of the class imbalance problem in convolutional
		neural networks}.
	\newblock {\em Neural Networks}, 106:249 -- 259.
	
	\bibitem[\protect\astroncite{Cameron et~al.}{2011}]{cameron2011robust}
	Cameron, A.~C., Gelbach, J.~B., and Miller, D.~L. (2011).
	\newblock {Robust Inference With Multiway Clustering}.
	\newblock {\em Journal of Business \& Economic Statistics}, 29(2):238--249.
	
	\bibitem[\protect\astroncite{Fern{\'a}ndez et~al.}{2004}]{fernandez2004mothers}
	Fern{\'a}ndez, R., Fogli, A., and Olivetti, C. (2004).
	\newblock {Mothers and Sons: Preference Formation and Female Labor Force
		Dynamics}.
	\newblock {\em The Quarterly Journal of Economics}, 119(4):1249--1299.
	
	\bibitem[\protect\astroncite{Fortin}{2015}]{fortin2015gender}
	Fortin, N.~M. (2015).
	\newblock Gender role attitudes and women's labor market participation:
	Opting-out, aids, and the persistent appeal of housewifery.
	\newblock {\em Annals of Economics and Statistics}, (117/118):379--401.
	\newblock Pre-print accessed.
	
	\bibitem[\protect\astroncite{Fryer and Levitt}{2010}]{fryer2010empirical}
	Fryer, R.~G. and Levitt, S.~D. (2010).
	\newblock {An Empirical Analysis of the Gender Gap in Mathematics}.
	\newblock {\em American Economic Journal: Applied Economics}, 2(2):210--40.
	
	\bibitem[\protect\astroncite{Goldin}{2006}]{goldin2006quiet}
	Goldin, C. (2006).
	\newblock {The Quiet Revolution that Transformed Women's Employment, Education,
		and Family}.
	\newblock {\em American Economic Review}, 96(2):1--21.
	
	\bibitem[\protect\astroncite{{Google's Ngram Viewer}}{}]{NgramViewer}
	{Google's Ngram Viewer}.
	\newblock \url{https://books.google.com/ngrams}.
	\newblock Accessed: 2018-07-30.
	
	\bibitem[\protect\astroncite{Guiso et~al.}{2008}]{guiso2008culture}
	Guiso, L., Monte, F., Sapienza, P., and Zingales, L. (2008).
	\newblock {Culture, Gender, and Math}.
	\newblock {\em Science}, 320(5880):1164--1165.
	
	\bibitem[\protect\astroncite{Hastie et~al.}{2009}]{hastie2009elements}
	Hastie, T., Tibshirani, R., and Friedman, J. (2009).
	\newblock {\em {The Elements of Statistical Learning: Data Mining, Inference,
			and Prediction}}.
	\newblock New York, NY: Springer.
	
	\bibitem[\protect\astroncite{Hornik}{1991}]{hornik1991approximation}
	Hornik, K. (1991).
	\newblock {Approximation Capabilities of Multilayer Feedforward Networks}.
	\newblock {\em Neural Networks}, 4(2):251--257.
	
	\bibitem[\protect\astroncite{{ISS}}{2016}]{ISS2016}
	{ISS} (2016).
	\newblock {Svenskarna och sociala medier 2016: En del av unders{\"o}kningen
		Svenskarna och internet 2016}.
	\newblock {Rapport from ISS}, Internetstiftelsen i Sverige.
	\newblock (ISS is a non-profit organization running the .se domain registry).
	
	\bibitem[\protect\astroncite{Leibbrandt and List}{2014}]{leibbrandt2014women}
	Leibbrandt, A. and List, J.~A. (2014).
	\newblock {Do Women Avoid Salary Negotiations? Evidence from a Large-scale
		Natural Field Experiment}.
	\newblock {\em Management Science}, 61(9):2016--2024.
	
	\bibitem[\protect\astroncite{Mazei et~al.}{2015}]{mazei2015meta}
	Mazei, J., H{\"u}ffmeier, J., Freund, P.~A., Stuhlmacher, A.~F., Bilke, L., and
	Hertel, G. (2015).
	\newblock {A Meta-analysis on Gender Differences in Negotiation Outcomes and
		Their Moderators}.
	\newblock {\em Psychological Bulletin}, 141(1):85.
	
	\bibitem[\protect\astroncite{Mikolov et~al.}{2013}]{mikolov2013efficient}
	Mikolov, T., Chen, K., Corrado, G., and Dean, J. (2013).
	\newblock {Efficient Estimation of Word Representations in Vector Space}.
	\newblock {\em ArXiv e-prints}.
	
	\bibitem[\protect\astroncite{Mueller and Plug}{2006}]{mueller2006estimating}
	Mueller, G. and Plug, E. (2006).
	\newblock {Estimating the Effect of Personality on Male and Female Earnings}.
	\newblock {\em ILR Review}, 60(1):3--22.
	
	\bibitem[\protect\astroncite{Mullainathan and
		Spiess}{2017}]{mullainathan2017machine}
	Mullainathan, S. and Spiess, J. (2017).
	\newblock {Machine Learning: An Applied Econometric Approach}.
	\newblock {\em Journal of Economic Perspectives}, 31(2):87--106.
	
	\bibitem[\protect\astroncite{Nollenberger et~al.}{2016}]{nollenberger2016math}
	Nollenberger, N., Rodr{\'\i}guez-Planas, N., and Sevilla, A. (2016).
	\newblock {The Math Gender Gap: The Role of Culture}.
	\newblock {\em American Economic Review}, 106(5):257--61.
	
	\bibitem[\protect\astroncite{North}{1990}]{north1990institutions}
	North, D. (1990).
	\newblock {\em {Institutions, Institutional Change and Economic Performance}}.
	\newblock Cambridge University Press.
	
	\bibitem[\protect\astroncite{Pope and Sydnor}{2010}]{pope2010geographic}
	Pope, D.~G. and Sydnor, J.~R. (2010).
	\newblock {Geographic Variation in the Gender Differences in Test Scores}.
	\newblock {\em Journal of Economic Perspectives}, 24(2):95--108.
	
	\bibitem[\protect\astroncite{Poria et~al.}{2017}]{poria2017adeeper}
	Poria, S., Cambria, E., Hazarika, D., and Vij, P. (2017).
	\newblock {A Deeper Look into Sarcastic Tweets Using Deep Convolutional Neural
		Networks}.
	\newblock {\em ArXiv e-prints}.
	
	\bibitem[\protect\astroncite{Rennie et~al.}{2003}]{rennie2003tackling}
	Rennie, J.~D., Shih, L., Teevan, J., and Karger, D.~R. (2003).
	\newblock {Tackling the Poor Assumptions of Naive Bayes Text Classifiers}.
	\newblock In {\em Proceedings of the 20th international conference on machine
		learning (icml-03)}, pages 616--623.
	
	\bibitem[\protect\astroncite{Shen et~al.}{2015}]{shen2015learning}
	Shen, L., Lin, Z., and Huang, Q. (2015).
	\newblock {Learning Deep Convolutional Neural Networks for Places2 Scene
		Recognition}.
	\newblock {\em ArXiv e-prints}.
	
	\bibitem[\protect\astroncite{Spencer et~al.}{1999}]{spencer1999stereotype}
	Spencer, S.~J., Steele, C.~M., and Quinn, D.~M. (1999).
	\newblock {Stereotype Threat and Women's Math Performance}.
	\newblock {\em Journal of Experimental Social Psychology}, 35(1):4--28.
	
	\bibitem[\protect\astroncite{{Statistics Sweden}}{2016a}]{scb2016females}
	{Statistics Sweden} (2016a).
	\newblock \url{http://www.statistikdatabasen.scb.se/sq/54996}.
	\newblock Data on female first names for 2016.
	
	\bibitem[\protect\astroncite{{Statistics Sweden}}{2016b}]{scb2016males}
	{Statistics Sweden} (2016b).
	\newblock \url{http://www.statistikdatabasen.scb.se/sq/54997}.
	\newblock Data on male first names for 2016.
	
	\bibitem[\protect\astroncite{Steinert-Threlkeld}{2018}]{steinert_threlkeld_2018}
	Steinert-Threlkeld, Z.~C. (2018).
	\newblock {\em Twitter as Data}.
	\newblock Elements in Quantitative and Computational Methods for the Social
	Sciences. Cambridge University Press.
	
	\bibitem[\protect\astroncite{Williamson}{2000}]{williamson2000new}
	Williamson, O.~E. (2000).
	\newblock {The New Institutional Economics: Taking Stock, Looking Ahead}.
	\newblock {\em Journal of Economic Literature}, 38(3):595--613.
	
	\bibitem[\protect\astroncite{Wu}{2018}]{wu2017gendered}
	Wu, A.~H. (2018).
	\newblock {Gendered Language on the Economics Job Market Rumors Forum}.
	\newblock {\em AEA Papers and Proceedings}, 108:175--179.
	
	\bibitem[\protect\astroncite{{Young} et~al.}{2017}]{young2017recent}
	{Young}, T., {Hazarika}, D., {Poria}, S., and {Cambria}, E. (2017).
	\newblock {Recent Trends in Deep Learning Based Natural Language Processing}.
	\newblock {\em ArXiv e-prints}.
	
\end{thebibliography}

\clearpage 

\setcounter{section}{0} 
\renewcommand{\thesection}{A\arabic{section}} 
\setcounter{subsection}{0} 
\renewcommand{\thesubsection}{A\arabic{subsection}} 
\setcounter{footnote}{0}
\setcounter{equation}{0}
\setcounter{figure}{0} 
\renewcommand{\thefigure}{A\arabic{figure}} 
\setcounter{table}{0} 
\renewcommand{\thetable}{A\arabic{table}}


\appendix
\section*{Appendix A. Model specifications and selection \label{paper3:appendix}}
\addcontentsline{toc}{section}{Appendix A. Model specifications and selection}

\subsection{Overview}

This appendix covers how natural language processing (NPL)- type of deep learning models are trained to detect gendered language in a unsupervised fashion. The models learn to predict which of the words \textit{he} and\textit{ she} that blank spots in Swedish tweets, which originally contained the words, should be filled with. One dimension of gendered language is used to infer the rest of the dimensions. This appendix covers specifications and model selection. The final model selected is evaluated on test data in the main article and, in addition, used as a measurement tool to evaluate the \#Metoo movement on another test data set. A ``control'' models is also trained to predict \textit{I} and \textit{we} in an equivalent manner. The ``control'' model is used to generate a comparison series for usage in the evaluation of the  \#Metoo movement.

\subsection{Data}

I have downloaded 58 million non-retweeted Swedish tweets from 2016-05-01 to 2017-04-30 (called Year 1 in the main article) from Twitter's API. Word embeddings are pretrained on the full dataset, but for the model training are two main sets are created. The he/she-set contains tweets with either the word \textit{he} or \textit{she} appearing and the I/we-set contains tweets with either the word \textit{I} or \textit{we} appearing. 

\begin{table}[H]
	\begin{centering}
		\noindent\resizebox{0.6\textwidth}{!}{
			\begin{tabular}{lcclcc}
				\hline
				& \multicolumn{2}{c}{Original} &  & \multicolumn{2}{c}{Adjusted}   \\ \cline{2-3} \cline{5-6}
				Set    & Count      & Class imbalance &  & Count      & Count 10-25 words \\
				& (millions) & (fraction)      &  & (millions) & (millions)        \\ \hline
				He/she & 1.5        & 0.255           &  & 1.5        & 1.1               \\
				I/we   & 8.3        & 0.240           &  & 1.5        & 1.1               \\ \hline
			\end{tabular}
		}
	\end{centering}
	
	\caption{\label{tab:Sets and sample sizes}Sets and sample sizes}
\end{table}

Table \ref{tab:Sets and sample sizes} (Original, Count) presents that the total count of the I/we-tweets are larger at 8.3 million than the he/she-tweets at 1.5 million. Since the I/we-tweets yields a ``control'' model, who’s performance is dependent on the size of the input dataset, the I/we-tweets are sub-sampled to have the same sample size as the he/she-tweets, e.g. 1.5 millions (Adjusted, Count). Also, table \ref{tab:Sets and sample sizes} (Original, Class imbalance) presents the class imbalance as the percent of appearances of the infrequent class, \textit{she} or \textit{we}. There are 25.5 percent she-tweets in the He/she-set and 24 percent we-tweets in the I/we-set. The two sets are similarly imbalanced, and, hence, no adjustment is needed for the I/we-tweets to yield a good ``control'' model in this regard. 

Table \ref{tab:Sets and sample sizes} (Adjusted, Count 10-25 words) displays the sample sizes of 1.1 million for both sets after they are cut to only included tweets with a word count between 10 and 25 words. The lower bound of 10 words is motivated by tweets below containing little information to predict on and reduces the original sample by around 10 percent. The upper bound is motivated by model technical reasons. When the input tweets are of similar length a better performing model is easier to find. The upper bound reduced the original sample by around 20 percent. An alternative would be to train multiple sub-models and thereby take into account input tweets of various length, but as the main goal is to detect gendered language a model using 10-25 word model is deemed sufficient.\footnote{The two sets are still similarly imbalanced at 25.1 and 23.9 percent  respectively.} Finally, the two sets are further subdivided by a 64-16-20 random split to a training, validation and test set each. 

One can expect noise in the binary variable indicating if \textit{he} or \textit{she} should be put in place of the placeholder. In Swedish dialects \textit{he} (han) can be substituted for \textit{him} (honom), e.g. the “grammatically correct” sentence \textit{I gave the ball to him} (Jag gav bollen till honom) can be transformed to \textit{I gave the ball to he} (Jag gav bollen till han). The expectation is due to language-historical reasons. I find that 'grammatically incorrect' Swedish dialects only introduces label noise at around 2 percent in the \textit{he-} class.\footnote{To investigate the concern of substitution due to “grammatically incorrect” Swedish dialects I evaluated a randomized balanced sample of 1000 he/she-tweets. The substitution occurs in he-tweets at a rate of 1.1 to 3.9 percent, with a point estimate of 2.2 percent. (From an exact binomial test where the null hypothesis is that the proportion is equal to zero).  There is no language-historical reason to expect the same kind of substitution for the female pronouns \textit{she} (hon) and \textit{her} (henne) and no substitution was found.} This level of noise is not a concern for the deep type of networks considered in this appendix.

\subsection{Preprocessing}
The tweets are preprocessed by replacing all user names, urls and hashtags with corresponding placeholders. \textit{NLTK Twitter Tokenizer} is used to divide the tweets into words. It is chosen since it preserve emoticons, such as ``{\smiley}'',  that oftentimes not is considered to be words. 

\subparagraph{Word embeddings}

A word vector (word embedding) is a $d$-dimensional vector striving to represent the words relationship with other words and it is used as input to the networks. Available word embeddings for Swedish are exclusively trained on Wikipedia text which most likely do not include the same type of language as Twitter, and hence have I trained my own. The word embeddings are trained with the Skip Gram algorithm of \citet{mikolov2013efficient} since it is a standard training method. They are trained on the full 58 million-tweet data with \textit{Gensim}. Implementation details are presented in table \ref{tab:Word-embeddings-implementation}. There are 575 000 tokens and 2.4 billion tokens to train on after the window length is set to ten. The words with a count lower than ten are replaced with a placeholder to be able to initialize such words during training. Swedish do not use compound nouns to the same degree as English, e.i. a ``full moon'' would be written ``fullmoon'' in Swedish (fullmåne), meaning that the model will have more word vectors but less co-references to solve than the standard English case.

\begin{table}[H]
	\begin{centering}
		\noindent\resizebox{0.45\textwidth}{!}{
			\begin{tabular}{lc}
				\hline
				Hyperparameter                           & Value        \\ \hline \hline
				Word vector dimmension                   & 300          \\
				Window Lenght                            & 10           \\
				Negative sample size                     & 5            \\
				Min Count                                & 10           \\
				Nr of epochs                             & 5            \\
				&              \\
				Feature                                  & Counts       \\ \hline \hline
				Sentences                                & 58 millions  \\
				Tokens                                   & 3.4 billions \\
				Tokens \textgreater Window Length        & 2.4 billions \\
				Unique tokens \textgreater Window Length & 575 000      \\ \hline
			\end{tabular}
		}
	\end{centering}	
	\caption{\label{tab:Word-embeddings-implementation}Word embedding implementation details }
\end{table}

\subparagraph{Masking of gendered words}

It can be argued that the model should capture more nuanced gender-colored language than simply using words such as \textit{boyfriend} and \textit{grandma} to predict. Therefore is in the main model 'too obviously' gender-colored words masked. In the masked data fictive words are created by collapsing the original ones. For example, the words \textit{mum} and \textit{dad} have been
masked to \textit{mumdad}. When using word vectors the new masked vectors are created by summing the original ones and re-normalizing them. I hand-select suitable words to mask by investigating lists of high predicting words, firstly, from the optimized Naive Bayes model and, later, from a LSTM neural network model trained on already masked data. I deem it suitable to mask the following types of words: 
\begin{enumerate}
	\item Tuples that consist of mappings of female words to a male equivalents, such as sister-brother. The set was expended with similar words as defined by the trained word vectors. Some high predicting words or neighbors was deemed impossible to map, such as witch (\textit{häxa}) and shallow-and-stupid-female (\textit{bimbo}), and are not masked for this reason.
	\item Family names of famous persons. 
	\item Personal names that consists of all Swedish resident's first names from a list obtained from \citet{scb2016females, scb2016males}, except of names that also is another word in Swedish (\textit{kalla, du}) or English (\textit{star, honey}). 
\end{enumerate}
Table \ref{tab:count-of-masked-words} displays the counts of the mask lists and the effective words found in training.\footnote{See online appendix table "Words used for masking" for complete lists of words used for masking.} The word count of the training set is reduced by three percent when masking.

\begin{table}[H]
	\begin{centering}
		\noindent\resizebox{0.8\textwidth}{!}{%
			\begin{threeparttable}
				\begin{tabular}{lllccc}
					\hline
					& Description         & Source & Tuple count & Word count & Word count found \\
					&                     &                            & on list     & on list    & in training      \\ \hline
					1 & high- predictors    & own                        & 421         & 932        & 753              \\
					2 & famous family names & own                        & 2           & 815        & 815              \\
					3 & first names         & Statistics Sweden          & 2           & 22751      & 5531            \\ \hline
				\end{tabular}
				\begin{tablenotes}[flushleft]
					\item[]Note: The tuple count is not equivalent to the word count on the mask lists. For example, the words \textit{wife} and \textit{husband} is in the same tuple as \textit{hubby}. The effective word counts found in the training set is displayed as it differs from the theoretically possible words to mask from the lists.
				\end{tablenotes}
			\end{threeparttable}
		}
	\end{centering}	
	\caption{\label{tab:count-of-masked-words}Counts of masked words }	
\end{table}

\subsection{Model specifications}

Three main model types are evaluated; a Naive Bayes model, a one-layers LSTM model and a three-layers CNN model. The Naive Bayes model function as a baseline and do not take into account word order, e.g. no time dimension is modelled. Both LSTM- and CNN model has shown competitive results on NPL-tasks\footnote{For an overview of recent results see \citet{young2017recent}.}, and both model a time dimension, although with different flavours.

An input sample is a sentence represented by $\displaystyle \mX=\{\vx_1, \dots, \vx_t \}$ where $\displaystyle  \vx_t $ represents a word by a one-hot (dummy-variable) encoded vector. $\displaystyle \mX\in\R^{t \times v}$ where $\displaystyle t$ is the number of words in the sentence and $\displaystyle v$ the number of words in the vocabulary. The curse of dimensionality composed by the general problem formulation is overcomed differently dependent on the model type. Count-based model such as the Naive Bayes reduce the one-hot encoded word representation to a scalar of the count of the word, $\displaystyle \mX\in\R^v$. The LSTM and CNN models instead uses word vectors. With word vectors a one-hot encoded word representation is reduced to $d$-dimensional vector striving to represent the words relationship with other words,  $\vx_{t}^{word\ vector}=\mB^{Reduce}\vx_t^{one- hot}$ where the $\mB^{Reduce}$-matrix is defined by the pretrained word embeddings. $\displaystyle \mX\in\R^{t \times d}$ where $d$ represents the dimensionality of the word vectors. When retraining of the word vectors is allowed the parameters in $\mB^{Reduce}$ are reestimated and the model operates on the full $\displaystyle \mX\in\R^{t \times v}$-matrix. The tweets are prepossessed by substituting \textit{he} and \textit{she} by a common placeholder. When word vectors are used, a representation for the common placeholder is initialized by summing up the original ones and re-normalizing them.

The output $\displaystyle y$ is a binary indicator for class 1, the \textit{she}-class. As standard for a two-class problem models are trained to minimize cross-entrophy loss $-\sum_{i}y_{i}log(\hat{y_{i}})+(1-y_{i})log(1-\hat{y_{i}})$,
where $\hat{y}=P(C_{1}|\mX)=\sigma(a)=1/1+e^{(-a)}$ with $a=log\left[P(\mX|C_{1})P(C_{1})/P(\mX|C_{2})P(C_{2})\right]$.
Minimizing cross-entrophy loss corresponds to maximizing the log-likelihood function behind a logistic model.

\subparagraph{(1) NB}
A baseline neural network Naive Bayes model is trained, which is equivalent to estimating a linear logistic regression with the count of each word as independent variables:

\begin{equation}
a=log\left[P(C_{1})/P(C_{2})\right]+\sum_{v=1}^{V}log\left[p_{1,v}/p_{2,v}\right]x_{v}=c+\sum_{v=1}^{V}b_{v}x_{v}
\end{equation}

where $p_{c,v}$ represents the probability for word $v$ in class $c$ and $x_{v}$ represents the count of word $v$ in the sentence. The model is naive since it does not take word order into account in the dimensionality reduction, it assumes that the words in each class are independent. In some standard descriptions of a Naive Bayes model is the baseline count for each word considered to be a hyperparameter \citep{rennie2003tackling}, which instead is estimated by the bias node (the baseline log odds ratio $log\left[P(C_{1})/P(C_{2})\right],$ the constant) in the logistic regression- specification.

\subparagraph*{(2) BoW- WV}
A naive 'bag of words'- specification on the word vectors is trained, which is equivalent to a logistic regression where each word vector dimension is an independent variable:

\begin{equation}
a=c+\sum_{t=1}^{T}\sum_{d=1}^{D}b_{d,t}x_{d,t}=f(\mX\mW+c)
\end{equation}

where $x_{t,d}$ represents the $d$:th dimension in word vector $t$ in the sentence. The right hand side- expression follow the standard of calling parameters in a network weights and representing them by $W$. $f$ is a linear activation function, e.g. $f(z)=z$. \citet{BolukbasietAl2016}  shows that unwanted gendered language is linearly separable in the word vectors trained on English Google News. This suggests that a linear specification might be sufficient and the current specification illustrate how performance change as compared to the non-linear LSTM and CNN specifications. In addition, it illustrate possible performance gains due to the usage of word vectors as compared to the Naive Bayes model.

\subparagraph*{(3) LSTM}
The LSTM network implemented in this paper has one layer and a fairly bare- bone structure compared to how LSTM nodes, or other recurrent nodes, are used for npl-tasks in the literature. However, for example, including a bidirectional structure did not aid for this specific task when tried during undocumented initial model selection. Thus, instead of trying various design tweaks, I decided to implement a CNN network as a comparison since it is a more flexible design.

A Long Short-Term Memory (LSTM) network is a subclass of Recurrent Neural Networks (RNN). For npl-tasks are input to the network a sequence of word vectors $\mX=\{\vx_1, \dots, \vx_t \}$, $\displaystyle \mX\in\R^{t \times d}$. An LSTM node allow for modeling longer time sequences by "automatically detrending" the time series. Various gates are used to "filter" the input of the LSTM node $\vx_t$ to the output of the node $h_t$. More specifically is a forget gate $f$, and input gate $i$ and and an output gate $o$ used:

\begin{multline}
	i=\sigma(\vu^i\vx_t+w^ih_{t-1}+b_i) \\ 
	f=\sigma(\vu^f\vx_t+w^fh_{t-1}+b_f) \\
	o=\sigma(\vu^o\vx_t+w^oh_{t-1}+b_o) \\
\end{multline}

where the $\vu$:s are respective parameter vectors, the $w$:s are respective weights (parameters) and the $b$:s respective bias weights (constants). The logistic function $\sigma(z)$ maps the gates to $(0,1)$. The "filtering" specifies as follows:

\begin{multline}
	h_t=tanh(c_t)o \\
	c_t=c_{t-1}f+\tilde{c_t}i \\
	\tilde{c_t}=tanh(\vu^c\vx_t+w^ch_{t-1}+b_c) \\
\end{multline}

where $\vu^c$ is a parameter vector, $w^c$ a weight and $b_c$ a bias weight. The output of the LSTM node $h_t$ is achieved by filtering the internal memory of the structure $c_t$ through the output gate $o$. In turn, the internal memory $c_t$ is a combination of its value at the previous time step $c_{t-1}$ filtered by the forget gate $f$ and a candidate update value $\tilde{c}$ filtered by the input gate $i$. In the proceeding step, the candidate value $\tilde{c}$ is a combination of the input $\vx_t$ and the output of the node at a previous timestep $h_{t-1}$. The $tahn$-functions maps to (-1,1) and thus allow for outputting or updating the sequence between the various time-steps. The full LSTM network consists of many LSTM nodes and specifies as follows:

\begin{equation}
\hat{y}=\sigma(a)=\sigma(\mW^hf(\vh_{t=T})+c^h)
\end{equation}

where $f$ is a rectified linear activation function (ReLu), $f(z)=max(z,0)$, applied element-wise. In contrast to the above models, $a$ is now a weighted sum of the output of the various LSTM nodes $\vh_{t=T}$ taken at the last timestep $T$. The full equation is equivalent to the last layer of the network. Dropout is applied to the last layer to regularize it. The features $\vh_{t=T}$ learned in the proceeding hidden LSTM layer are randomly dropped out with probability p during training and the parameters $\mW^h$ rescaled with the same probability when training is finished. Both the number of nodes in the LSTM layer and the fraction of dropout are examples of hyperparameters that can be tuned.

\subparagraph*{(4) CNN}
A three-convolutional layer CNN model is trained which step-wise reduce the number of n-grams from $\{4,5\}$ in the first convolutional layer, to 3 in the second and 2 in the third. Each convolutional layer consists of 100 kernels. The input layer is zero-padded and the convolutions are narrow/valid. Max pooling is performed with a window size of 2 and, as conventional, the same stride. The last fully connected layer consists of 100 neurons. The CNN design is similar to the one found in \citet{poria2017adeeper}, but overall it follows a standard CNN architecture for a NPL-task. Instead of operating on an image, a NPL CNN operates on an input sentence represented by the matrix $\displaystyle \mX\in\R^{t \times d}$ with 1-d convolutions. Each kernel is applied to a window of h words to produce a new feature.

\begin{figure}[t!]
	\centering
	\includegraphics[width=0.6\textwidth]{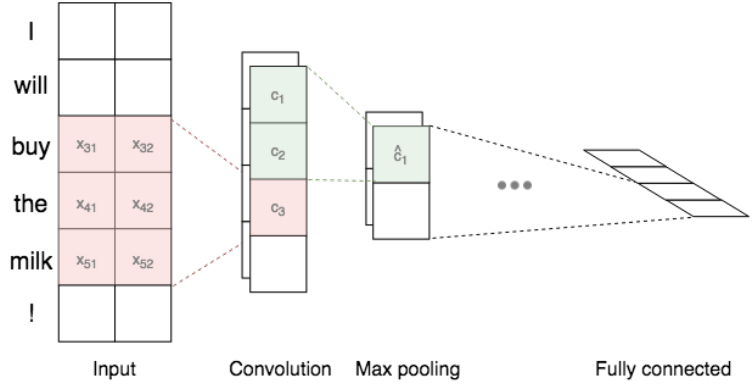}
	\caption{The figure illustrates a CNN structure with an example of the input sentence $\mX=\{\vx_1, \dots, \vx_t \}=\{I, will, \dots, ! \}$ where t=6 and d=2. In the figure is a kernel  $\vk=\{k_1, \dots, k_6 \}$ set to h=3 illustrated sliding over the window $\vx_{3:5}=\{x_{3,1},x_{3,2},x_{4,1}, \dots, x_{5,2} \}$ to produce the new feature $c_3$ in the new feature map $\vc$. The figure also illustrate the max pooling- step with a kernel size of h=2 with $\hat{c_1}=max\{\{c_{1},c_{2}\}\}$. That multiple kernels are applied in each set is illustrated by having 2 kernels and feature maps. The possibility of having multiple convolutional layers is in figure 1 illustrated by the three horizontal dots ($\dots$). The figure also illustrate that the last fully connected layer in the CNN is obtained by concatenating the last max pooling layer. (Dropout is not illustrated.)}
	\label{fig:cnn_illustartion}
\end{figure}

In a CNN model a convolutional layer is normally built up in two steps: a convolution and max pooling. The model structure is illustrated in \figref{fig:cnn_illustartion}. In the convolutional step, a kernel $\vk\in\R^{hd}$ is applied to a window $\vx_{t:t+h}\in\R^{hd}$, a concatenation of vectors from $\vx_t$ to $\vx_{t+h}$, to produce a new feature $c_i=f(\vk^{\rm T}\vx_{t:t+h-1}+b)$. $f$ is a rectified linear activation function (ReLu), $f(z)=max(z,0)$.  One kernel is applied to all possible windows to produce a new feature map $\vc= \{c_1, \dots, c_{t-h+1} \}$. The parameters in the kernel is shared across the windows. In the max pooling step, with downsampling and conventional stride settings, is another filter applied, $\hat{\vc_j}=max\{\vc_{j:j+h-1}\}$, yielding a final feature vector $\vz$.  In each convolutional layer is normally multiple kernels and feature maps applied. A CNN model can include many convolutional layers after each other. The last layer in the CNN is a fully connected layer with concatenated input $\vz$ from the last max-pooling layer, $a=\mW\vz+c$ (with the set-up used in this paper). Dropout is again applied to the last layer, but now the features in the last max-pooling layer randomly are dropped out. 

\subsection{Hyperparameter tuning}

Hyperparameters are tuned on the masked he/she-tweets and, thus, model selection is optimized based on finding a model which takes into account more nuanced gendered-color language. An increase in accuracy of 1 percentage point is considered an improvement.\footnote{The limit is based on eyeballing what \citet{young2017recent} consider to be marked improvements in their overview of literature on NLP in deep learning.} 

The main set of hyperparameters tuned is if word vectors should be retrained and methods to deal with class imbalance. Re-train of the word vectors are only applicable for the LSTM and CNN models. Depending on the quality of the initialized word embeddings for the task at hand, retraining can decreased performance by adjusting the word vectors in the smaller training set only to make them less generalizable.\footnote{To clarify, only words found in the training set are included in the model.}  I also investigate methods to deal with class imbalanced during training. Undocumented initial model selection showed that overall accuracy decreases without such a method. The first method investigated consists of weightening the loss function to pay more attention to the underrepresented class. There exist around three times more he-tweets than she-tweets in the training data, and the she-class is thus upweighted by a factor close to three. The method is standard, but since binary cross entrophy loss is used errors will tend towards infinity, and the weightening might be less effective. The second method investigated consists of making each mini batch balanced. The strategy is mentioned by \citet{buda2108asystematic} and used in \citet{shen2015learning}.  The balancing of each mini batch is done by randomly redrawing samples until a sufficient number of the underrepresented class is available and then randomly cutting away samples representing the overrepresented class. Each mini batch approximate the empirical distribution of how the data would be if it was balanced. The method can be thought of as a stochastic gradient decent- version of more standard methods of oversampling or undersampling. 

After the main set of hyperparameters have been optimized, the following are also tuned: For the LSTM network the number of nodes and the fraction of dropout applied to the weights are investigated, since they are important hyperparameters guiding generalization performance. More specifically, $\{125,250, 500\}$ nodes and $\{0,0.25, 0.5\}$ in dropout fraction was tried. For the CNN network are the fraction of dropout tuned, again investigating fractions of $\{0,0.25, 0.5\}$. The numbers of filters are not optimize since 100 filters per layer seems to be fairly standard.

The hyperparameters for the other models used in the main article, the unmasked he/she-model and the unmasked I/we-model, are set to be equivalent to the ones found for the masked he/she-model as they consider a similar problem. In addition, not all hyperparamters are tuned for the masked he/she-model since it would cost to much computational resources. Only normalized 300-dimensional word vectors trained by me have been tried. All tweets are prepocessed to the same input length by adding padding to the right of the original input vector, which is applicable only for all model types except the Naive Bayes model. All training is performed on the scale of one tenth (1/10:th) epoch since the sample sizes are large. The networks are trained with a mini batch size of 100 samples for max 5 epochs. Training stops earlier if the validation accuracy do not increase by 0.1 percentage point after 1 epochs. The untuned hyperparameters strive to follow standard settings for natural language processing- types of neural networks.

\subsection{Validation results}

\begin{table}[H]
	\begin{centering}
		\noindent\resizebox{0.7\textwidth}{!}{%
			\begin{threeparttable}
				\begin{tabular}{lcccc}
					\hline
					& (1)    & (2)            & (3)            & (4)         \\
					Model type                  & NB     & BoW- WV        & LSTM           & CNN         \\ \hline
					&        &                &                &             \\
					ROC AUC                     & 0.7315 & 0.7180         & 0.7640         & 0.7588      \\
					&        &                &                &             \\
					Accuracy                    & 0.7541 & 0.7567         & 0.7733         & 0.7721      \\
					&        &                &                &             \\
					Hyperparameters:            &        &                &                &             \\
					Training w. class imbalance &        & Balanced Batch & Balanced Batch & Weightening \\
					Training of word vectors    &        &                & No             & No          \\
					Nr of nodes                 &        &                & 125            & 100         \\
					Dropout fraction            &        &                & 0.25           & 0.5        \\ \hline
				\end{tabular}
				\begin{tablenotes}[flushleft]
					\item[]Note: Hyperparameter optimized results for masked He/she-data for each model type evaluted on the unbalanced validation set.
				\end{tablenotes}
			\end{threeparttable}
		}
	\end{centering}	
	\caption{\label{tab:validation-summary}Validation summary results }	
\end{table}

Table \ref{tab:validation-summary} presents a summary of the hyperparameter-optimized results for each model type. The NB model presented in the first column is trained with \textit{SciKit's}  multinomial Naive Bayes implementation, training with stochastic gradient decent in the neural network framework showed worse performance (not shown). The NB model's accuracy at 0.754 precisely beat the simple classification rule of predicting any sample to the he-class, which yield an accuracy of 0.749. The second column show that a naive model with word vectors as input (BoW-WV) do not aid performance. The third column presents the best performing model, a LSTM neural network model which reach a ROC AUC-score of 0.764 and an accuracy of 0.773. The fourth column displays that the CNN is a close runner-up at a ROC AUC-score of 0.758 and an accuracy of 0.772. Thus, introspection of table  \ref{tab:validation-summary} shows that the major performance gains arise when taking word order into account by adding non-linearities in the LSTM- and CNN- model types. More precisely, in the LSTM model the ROC AUC-score is 0.03 points higher and accuracy is 0.02 points higher than in the NB model. More nuanced test set result for the Naive Bayes and the LSTM model is found in the main article, but the ROC AUC-scores are 0.7372 and 0.7629 respectively.

Table \ref{tab:hyperparameter-opt-1} displays result for the main set of hyperparameters iterated over, if word vectors should be retrained and which method that should be used to deal with the class imbalance. The table shows that regardless of the which method that is used to deal with the class imbalance, re-training of the word vector hurts performance. One likely reason is that the word vectors are trained on the same type of data and thus retraining only decrease generalization performance. Table \ref{tab:hyperparameter-opt-1} also shows that higher performance is achieved, regardless of retraining of word vectors or not, for the LSTM model type when each mini batch is balanced (Balanced Batch) instead of when the loss function is weighted (Weightening). For the CNN model type is the result less clear, but by favoring ROC AUC as the evaluation metric over accuracy one reaches the conclusion that weightening is performance enhancing. For the BoW-WV model was the discrepancy between balancing each mini batch and weighting the loss function also not large (not shown).  One possible reason of why the discrepancy between the methods are larger for the LSTM model type can be that an LSTM node 'detrend' the input data, rendering it more sensitive to imbalanced data.

\begin{table}[H]
	\begin{centering}
		\noindent\resizebox{0.7\textwidth}{!}{%
			\begin{threeparttable}
				\begin{tabular}{lcccc}
					\hline
					&                &                &             &             \\ 
					LSTM                        &                &                &             &             \\ \hline
					&                &                &             &             \\
					ROC AUC                     & 0.7410         & 0.7540         & 0.6486      & 0.7061      \\
					Accuracy                    & 0.7672         & 0.7698         & 0.7664      & 0.7651      \\
					&                &                &             &             \\
					Training of word vectors    & Yes            & No             & Yes         & No          \\
					Training w. class imbalance & Balanced Batch & Balanced Batch & Weightening & Weightening \\
					&                &                &             &             \\
					CNN                         &                &                &             &             \\ \hline
					&                &                &             &             \\
					ROC AUC                     & 0.7331         & 0.7566         & 0.7325      & 0.7588      \\
					Accuracy                    & 0.7683         & 0.7725         & 0.7691      & 0.7721      \\
					&                &                &             &             \\
					Training of word vectors    & Yes            & No             & Yes         & No          \\
					Training w. class imbalance & Balanced Batch & Balanced Batch & Weightening & Weightening \\ \hline
				\end{tabular}
				\begin{tablenotes}[flushleft]
					\item[]Note: Iterations over hyperparameters for masked He/She-data for the LSTM- and CNN model type evaluated on the original unbalanced validation set. The LSTM model have 250 nodes and a dropout fraction of 0.5. The CNN have 100 filters and a dropout fraction of 0.5.
				\end{tablenotes}
			\end{threeparttable}
		}
	\end{centering}	
	\caption{\label{tab:hyperparameter-opt-1} Main hyperparameter optimization result}	
\end{table}

\subsection{Further suggestions}
Some interesting findings of this project is beyond the scope of the current purpose. The idea of capturing gendered language can be used in a more practical application. Then, it would be interesting to investigate how a non-linear neural network model performs when the scalar word count representation is used as input instead of a word vector representation. It a smaller model to implement and if yielding the same performance it would possibly allow for a memory- or speed-enhancing experience by an end user. However, to my knowledge it is not much discussed in the deep npl literature. Also, it would be interesting to delve deeper into how differently the models predict, beyond using a simple metric such as ROC AUC-scores or accuracy. An undocumented finding of this project is that the models predict individual tweets very differently, the overlap of how they capture gendered language is not huge. If used in a practical application, one would like to favor a model who mark words and phrases as gendered that agree with the thoughts of the end users on the matter.

\subsection{Summary}
Three model types are tried: a Naive Bayes, a LSTM and a CNN model. The best performing model type is an 1-layer LSTM model with 125 nodes and a dropout fraction of 0.25 which is trained by balancing each mini batch and not allowing for retraining of the word vectors.
\clearpage

\setcounter{section}{0} 
\renewcommand{\thesection}{B\arabic{section}} 
\setcounter{subsection}{0} 
\renewcommand{\thesubsection}{B\arabic{subsection}} 
\setcounter{footnote}{0}
\setcounter{equation}{0}
\setcounter{figure}{0} 
\renewcommand{\thefigure}{B\arabic{figure}} 
\setcounter{table}{0} 
\renewcommand{\thetable}{B\arabic{table}}


\section*{Appendix B. Additional Information and Results \label{paper3:resultappendix}}
\addcontentsline{toc}{section}{Appendix B. Additional Information and Results}

\subsection*{B1 List of survey questions used to capture gender norms \label{tab:Survey questions used to capture gender norms}}
\addcontentsline{toc}{subsection}{List of survey questions used to capture gender norms}

\begin{itemize}
	\item World Values Surveys
	\begin{enumerate}
		\item When jobs are scarce, men should have more right to a job than women.
		\item On the whole, men make better political leaders than women do.
		\item Most men are better suited emotionally for politics than are most women.
	\end{enumerate}
	\item General Social Survey (US)
	\begin{enumerate}
		\item Most men are better suited emotionally for politics than are most women.
		\item A working mother can establish just as warm and secure a relationship with her children as a mother who does not work.
		\item A preschool child is likely to suffer if his or her mother works.
		\item It is much better for everyone involved if the man is the achiever outside the home and the woman takes care of the home and family.
	\end{enumerate}
	\item National Longitudinal Survey of 1972 (US)
	\begin{enumerate}
		\item A working mother of pre-school children can be just as good a mother as the woman who doesn’t work.
		\item High schools counselors should urge young women to train for jobs which are now held mainly by men.
		\item It is more important for a wife to help her husband than to have a career herself.
		\item It is usually better for everyone involved if the man is the acheiver outside the home and the woman takes care of the home and family.
		\item Many qualified women can’t get good jobs; men with the same skills have much less trouble.
		\item Men should be given first chance at most jobs because they have the primary responsibility for providing for a family.
		\item Most women are happiest when they are making a home and caring for children.
		\item Most women are just not interested in having big and important jobs.
		\item Schools teach women to want the less important jobs.
		\item Young men should be encouraged to take jobs that are usual filled by women (nursing, secretarial, work. etc.).
	\end{enumerate}	
\end{itemize}

\subsection*{B2 Tables and graphs}
\addcontentsline{toc}{subsection}{Tables and graphs}

\begin{table}[H]
	\begin{centering}
		\noindent\resizebox{0.45\textwidth}{!}{%
			\begin{threeparttable}
				\begin{tabular}{llll}
					\hline \hline
					Year 1            & Year 2            & Year 2            & Year 2            \\ \hline
					2016-09-05 & 2017-05-07 & 2017-09-24 & 2018-03-17 \\
					2016-09-19 & 2017-05-25 & 2017-10-03 & 2018-04-17 \\
					2016-10-15 & 2017-05-27 & 2017-10-09 & 2018-04-18 \\
					2016-10-16 & 2017-06-11 & 2017-11-12 & 2018-04-22 \\
					2016-10-18 & 2017-06-12 & 2017-11-14 &   \\
					2016-10-20 & 2017-06-13 & 2017-11-17 &   \\
					2016-10-29 & 2017-06-14 & 2017-11-18 &   \\
					2016-11-01 & 2017-06-15 & 2017-11-19 &   \\
					2016-11-02 & 2017-06-16 & 2017-11-20 &   \\
					2016-11-03 & 2017-06-17 & 2017-11-21 &   \\
					2016-11-04 & 2017-06-18 & 2017-11-22 &   \\
					2016-11-05 & 2017-06-20 & 2017-11-23 &   \\
					2016-11-28 & 2017-06-25 & 2017-12-01 &   \\
					2017-02-15 & 2017-07-05 & 2017-12-09 &   \\
					2017-02-26 & 2017-07-23 & 2017-12-27 &   \\
					2017-03-11 & 2017-07-30 & 2018-01-09 &   \\
					2017-03-12 & 2017-07-31 & 2018-01-11 &   \\
					2017-03-13 & 2017-08-01 & 2018-01-28 &   \\
					2017-03-14 & 2017-08-14 & 2018-01-29 &   \\
					2017-03-18 & 2017-08-29 & 2018-01-30 &   \\
					2017-03-28 & 2017-09-01 & 2018-02-18 &   \\
					2017-04-14 & 2017-09-23 & 2018-02-20 &   \\ \hline \hline
				\end{tabular}
				\begin{tablenotes}[flushleft]
					\item[]Note: UTC dates are presented because it is the level where downloading
					failures happen. Swedish dates differ to UTC dates with +1 or +2 hours
					(depending on daylight saving time).
				\end{tablenotes}
			\end{threeparttable}
		}
	\end{centering}	
	\caption{\label{tab:Missing-UTC-dates}Missing UTC dates.}
\end{table}

\begin{table}[H]
	\begin{centering}
		\noindent\resizebox{1\textwidth}{!}{
			\begin{threeparttable}
				\begin{tabular}{lllccc}
					\hline \hline
					Corpus                & Description                    & Timespan  & She Frac & She \& Her Frac & n (thusands) \\ \hline
					Bloggmix              & Popular blogs                  & 2008-2017 & 0.429    & 0.441           & 2 713        \\
					Familjeliv            & Disussion forum on family life & 2004-2017 & 0.418    & 0.424           & 36 788       \\
					Bloggmix              & Popular blogs                  & 1998-2007 & 0.370    & 0.376           & 182          \\
					SOUs                  & Governmental reports           & 1995-2016 & 0.366    & 0.367           & 160          \\
					Göteborgsposten       & Newspaper                      & 2004-2013 & 0.336    & 0.338           & 1 292        \\
					Flashback             & Dicussion forum                & 2000-2017 & 0.327    & 0.351           & 17 029       \\
					Webnews               & Newspapers                     & 2004-2013 & 0.278    & 0.281           & 1 682        \\
					Forskning \& Framsteg & Popular science magasine       & 1992-1996 & 0.272    & 0.284           & 1.8          \\
					Wikipedia             &                                & 2017      & 0.225    & 0.231           & 1 050        \\ \hline \hline
				\end{tabular}
				\begin{tablenotes}[flushleft]
					\item[]Note: 
					
					She Frac: \textit{she}/(\textit{she}+\textit{he}) where \textit{she} is the count of the word \textit{she} (hon), inclusive of \textit{She} (Hon), and \textit{he} is the count of the word \textit{he} (han), inclusive of \textit{He} (Han). 
					
					She \& Her Frac: (\textit{she}+\textit{her}) /(\textit{she}+\textit{her}+\textit{he}+\textit{him}) where, in addition to the definitions above, \textit{her} is the count of the word \textit{her} (henne), inclusive of \textit{Her} (Henne) and \textit{him} is the count of the word \textit{him} (honom), inclusive of \textit{Him} (Honom).
					
					n (thousands) is the total count of the words \textit{he}, \textit{He}, \textit{she} and \textit{She} in thousands.
					
					The various text sources are retrieved from Språkbanken/Korp (2018).
				\end{tablenotes}
			\end{threeparttable}
		}
	\end{centering}	
	\caption{\label{tab:Class imbalance in other Swedish text sources}Class imbalance in other Swedish text sources}
\end{table}

\begin{small}
	\begin{center}
		\begin{longtable}{lll}
			\caption{\label{tab:highPredictingWords-balanced}Frequent words over the predicted probability distribution evaluated at a balanced sample} \\
			&                                        &                                \\ \hline \hline
			WC   & English                                         & Swedish                                  \\ \hline
			\endfirsthead
			\multicolumn{3}{c}%
			{\tablename\ \thetable\ -- \textit{Continued from previous page}} \\
			\hline \hline
			WC   & English                                         & Swedish                                  \\ \hline
			\endhead
			\hline \hline \multicolumn{3}{r}{\textit{Continued on next page}} \\
			\endfoot
			\hline \hline
			\multicolumn{3}{l}{Note: The most “male” word with a WC of close to zero is} \\
			\multicolumn{3}{l}{at the top of the table and the most “female” word with a} \\
			\multicolumn{3}{l}{WC close to one is at the bottom of the table. The table is} \\
			\multicolumn{3}{l}{generated in the following way: The median predicted} \\
			\multicolumn{3}{l}{probability for each word is calculated from the predicted} \\
			\multicolumn{3}{l}{probability of each tweet, named Word Color (WC). The WC of} \\
			\multicolumn{3}{l}{all words is binned into 20 groups and for each quantile is} \\
			\multicolumn{3}{l}{the 5 most frequent words displayed. The table is generated} \\
			\multicolumn{3}{l}{on the test set and evaluated by the masked neural network} \\
			\multicolumn{3}{l}{model. The translation from Swedish is made by the author.} \\
			\endlastfoot
			0.01 & contract                                        & kontrakt                                 \\
			0.02 & united                                          & united                                   \\
			0.03 & games                                           & matcher                                  \\
			0.03 & player                                          & spelare                                  \\
			0.04 & club                                            & klubb                                    \\
			0.05 & the season                                      & säsongen                                 \\
			0.06 & goal                                            & mål                                      \\
			0.07 & the game                                        & matchen                                  \\
			0.08 & game                                            & match                                    \\
			0.09 & play                                            & spela                                    \\
			0.10 & the boll                                        & bollen                                   \\
			0.11 & team                                            & lag                                      \\
			0.13 & played                                          & spelade                                  \\
			0.14 & football                                        & fotboll                                  \\
			0.15 & penalty                                         & straff                                   \\
			0.16 & field                                           & plan                                     \\
			0.18 & ready                                           & klar                                     \\
			0.18 & play                                            & spelar                                   \\
			0.18 & miss                                            & missar                                   \\
			0.20 & em                                              & em                                       \\
			0.21 & europe                                          & europa                                   \\
			0.23 & the chance                                      & chansen                                  \\
			0.23 & worse                                           & sämre                                    \\
			0.25 & president                                       & president                                \\
			0.25 & \textless{}familyname\textgreater{}             & \textless{}familyname\textgreater{}      \\
			0.27 & short                                           & kort                                     \\
			0.27 & leave                                           & lämnar                                   \\
			0.27 & last                                            & förra                                    \\
			0.28 & score                                           & poäng                                    \\
			0.30 & bad                                             & dålig                                    \\
			0.32 & smallest                                        & minst                                    \\
			0.32 & latest                                          & senaste                                  \\
			0.34 & manage                                          & lyckas                                   \\
			0.34 & still there                                     & kvar                                     \\
			0.35 & before                                          & före                                     \\
			0.35 & against                                         & mot                                      \\
			0.37 & were                                            & varit                                    \\
			0.39 & good                                            & bra                                      \\
			0.40 & fuck                                            & fan                                      \\
			0.40 & in                                              & in                                       \\
			0.43 & like                                            & ju                                       \\
			0.43 & come                                            & kommer                                   \\
			0.44 & in                                              & i                                        \\
			0.44 & had                                             & hade                                     \\
			0.44 & .                                               & .                                        \\
			0.45 & \textless{}user\textgreater{}                   & \textless{}user\textgreater{}            \\
			0.46 & ,                                               & ,                                        \\
			0.47 & \_\_\_                                          & \_\_\_                                   \\
			0.48 & are                                             & är                                       \\
			0.48 & at                                              & att                                      \\
			0.50 & \textless{}hashtag\textgreater{}                & \textless{}hashtag\textgreater{}         \\
			0.51 & and                                             & och                                      \\
			0.52 &                                                 &                                          \\
			0.53 & I                                               & jag                                      \\
			0.54 & \textless{}url\textgreater{}                    & \textless{}url\textgreater{}             \\
			0.55 & you                                             & dig                                      \\
			0.57 & love                                            & älskar                                   \\
			0.58 & me                                              & mig                                      \\
			0.59 & my                                              & mitt                                     \\
			0.59 & \textless{}boygirl\textgreater{}                & \textless{}killetjejkillle\textgreater{} \\
			0.61 & my                                              & mina                                     \\
			0.62 & *heart eyes*                                    & *heart eyes*                             \\
			0.63 & children                                        & barn                                     \\
			0.63 &                                                 &                                          \\
			0.64 & *heart*                                         & *heart*                                  \\
			0.66 & \textless{}womenmen\textgreater{}               & \textless{}kvinnormän\textgreater{}      \\
			0.66 & \textless{}thedaughtertheson\textgreater{}      & \textless{}dotternsonen\textgreater{}    \\
			0.66 & my                                              & min                                      \\
			0.67 & \textless{}brothersister\textgreater{}          & \textless{}brorsyster\textgreater{}      \\
			0.69 & \textless{}sondaugther\textgreater{}            & \textless{}dotterson\textgreater{}       \\
			0.71 & friend                                          & \textless{}väninnavän\textgreater{}      \\
			0.72 & \textless{}mum'sdad's\textgreater{}             & \textless{}mammaspappas\textgreater{}    \\
			0.72 & \textless{}grandfathergrandmother\textgreater{} & \textless{}farfarfarmor\textgreater{}    \\
			0.74 & beautiful                                       & vacker                                   \\
			0.75 & \textless{}mumdad\textgreater{}                 & \textless{}mammapappa\textgreater{}      \\
			0.76 & fi                                              & fi                                       \\
			0.77 & \textless{}thewomenthemen\textgreater{}         & \textless{}kvinnornamännen\textgreater{} \\
			0.77 & \textless{}grandfathergrandmother\textgreater{} & \textless{}morfarmormor\textgreater{}    \\
			0.78 & pink                                            & rosa                                     \\
			0.80 & yearly                                          & åriga                                    \\
			0.81 & makeup                                          & smink                                    \\
			0.81 & nails                                           & naglar                                   \\
			0.82 & hagen                                           & hagen                                    \\
			0.82 & the cookie                                      & kakan                                    \\
			0.82 & pippi                                           & pippi                                    \\
			0.86 & noora                                           & noora                                    \\
			0.87 & book                                            & book                                     \\
			0.87 & thatcher                                        & thatcher                                 \\
			0.88 & raped                                           & våldtagen                                \\
			0.90 & zara                                            & zara                                     \\
			0.90 & dress                                           & klänning                                 \\
			0.91 & pregnant                                        & gravid                                   \\
			0.92 & veil                                            & slöja                                    \\
			0.94 & the lady                                        & tanten                                   \\
			0.94 & \textless{}hishers\textgreater{}                & \textless{}hanshennes\textgreater{}      \\
			1.00 & lift                                            & lyft                                     \\
			1.00 & credit                                          & credit                                   \\
			1.00 & free                                            & free                                     \\
			1.00 & casino                                          & casino                                   \\
			1.00 & code                                            & code                                    \\
		\end{longtable}
	\end{center}
\end{small}

\begin{figure}[H]
	\begin{centering}
		\includegraphics[width=1\textwidth]{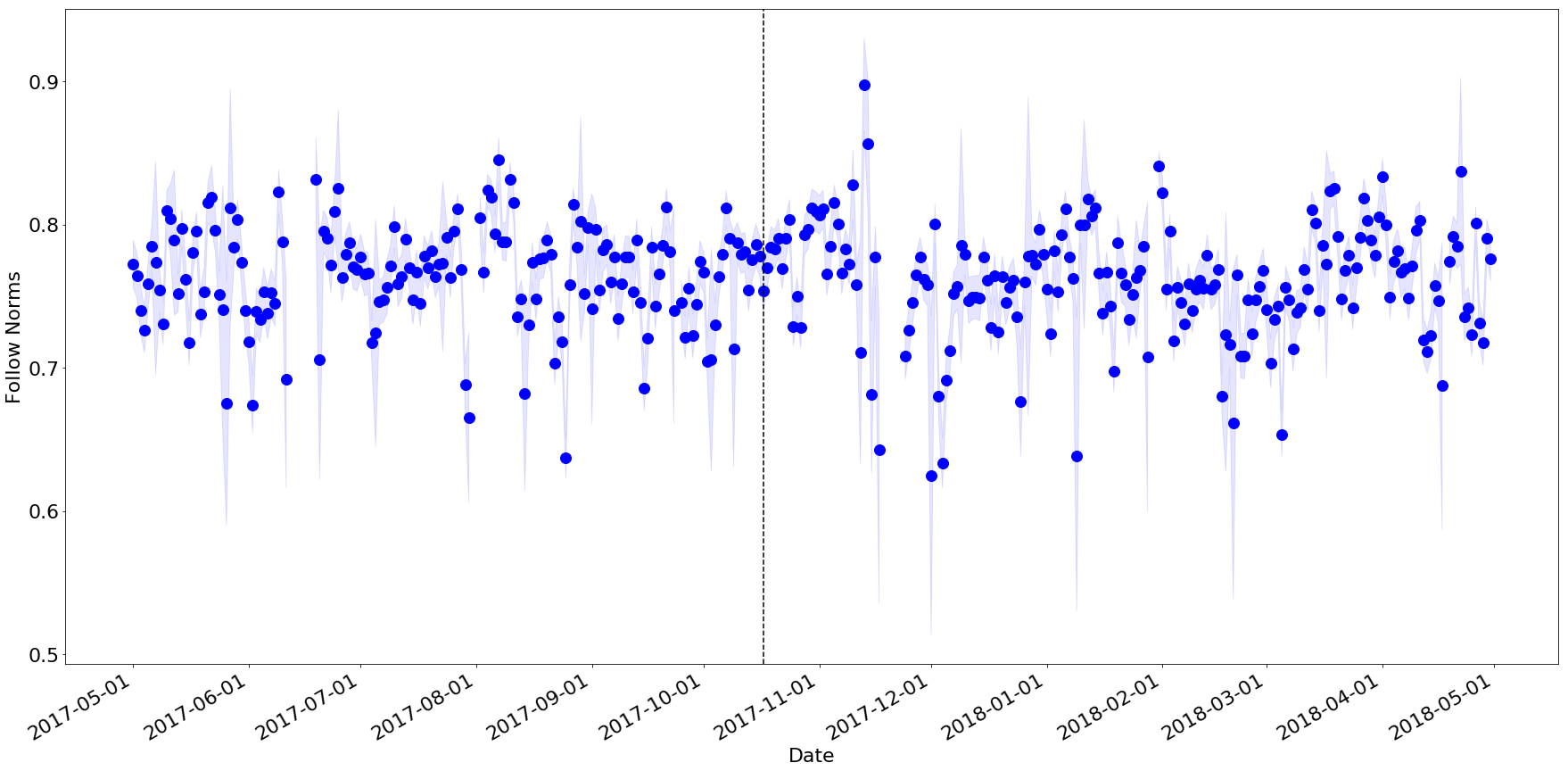}
		\par\end{centering}
	\caption{\label{fig:Gender norms over time}Gender norms over time}
	\begin{flushleft}
		\footnotesize{Note: The main dependent variable \textit{Follow Norms}, a binary indicator on if Geni predicts the tweet correct, is displayed on the y-axis. The evaluation set of Year 2 is used. The vertical dotted line represent the Metoo event at the 17th of October 2017 and the lighter shaded regions represent daily confidence intervals.}
	\end{flushleft}
\end{figure}

\begin{table}[H]
	\begin{centering}
		\noindent\resizebox{1\textwidth}{!}{%
			\begin{threeparttable}
				\begin{tabular}{lcccc}
					\hline \hline
					& Time Trends        & Time Trends        & Time Trends        & Time Trends      \\
					& Either Side        &                    & User FE            & User FE + Day FE \\
					& (1)                & (2)                & (3)                & (4)              \\
					Dep Var                           & Follow Norms * 100 & Follow Norms * 100 & Follow Norms * 100 &  Follow Norms * 100  \\ \hline
					&                    &                    &                    &                  \\
					Constant                          &                    & 76.842***          & 77.422***          &                  \\
					&                    & (0.738)            & (0.243)            &                  \\
					t                                 &                    & 0.020              & 0.006              &                  \\
					&                    & (0.021)            & (0.006)            &                  \\
					t\textasciicircum{}2              &                    & -0.000             & -0.000             &                  \\
					&                    & (0.000)            & (0.000)            &                  \\
					After Metoo                       & 88.290***          & 11.449**           & 19.684***          & 6.354            \\
					& (5.598)            & (5.646)            & (1.632)            & (7.770)          \\
					After Metoo*t                     & -0.086**           & -0.107**           & -0.154***          & -0.060           \\
					& (0.044)            & (0.048)            & (0.014)            & (0.066)          \\
					After Metoo*t\textasciicircum{}2  & 0.000*             & 0.000**            & 0.000***           & 0.000            \\
					& (0.000)            & (0.000)            & (0.000)            & (0.000)          \\
					Before Metoo                      & 76.842***          &                    &                    & -1.730           \\
					& (0.738)            &                    &                    & (1.234)          \\
					Before Metoo*t                    & 0.020              &                    &                    & 0.017            \\
					& (0.021)            &                    &                    & (0.033)          \\
					Before Metoo*t\textasciicircum{}2 & -0.000             &                    &                    & -0.000           \\
					& (0.000)            &                    &                    & (0.000)          \\
					&                    &                    &                    &                  \\
					Time FE                           & No                 & No                 & No                 & Yes              \\
					User FE                           & No                 & No                 & Yes                & Yes              \\
					SEs                               & Clustered on day   & Clustered on day   & Clustered on user  & Clustered on     \\
					&                    &                    &                    & day \& user      \\
					&                    &                    &                    &                  \\
					Tweets                            & 989 028            & 989 028            & 849 518            & 1 053 209        \\
					Days (T)                          & 319                & 319                & 319                & 365              \\
					Users (N)                         & 95 025             & 95 025             & 35 338             & 65 165           \\
					&                    &                    &                    &                  \\
					Data                              & Year2              & Year2              & Year2              & Year2/1          \\ \hline \hline
				\end{tabular}
				\begin{tablenotes}[flushleft]
					\item[]Note: The coefficients in the table can directly be interpreted as percentage points since the binary dependent variable has been multiplied by hundred. Aggregating to daily data in specification (1) and (2) and estimating Newey West standard errors with a lag length of 4 does not change the results. {*}{*}{*} p$<$0.01, {*}{*} p$<$0.05, {*} p$<$0.1.
				\end{tablenotes}
			\end{threeparttable}
		}
	\end{centering}	
	\caption{\label{tab:metoo-with-time-trends}Metoo results with time trends}
\end{table}

\begin{table}[H]
	\begin{centering}
		\noindent\resizebox{0.7\textwidth}{!}{%
			\begin{threeparttable}
				\begin{tabular}{lcccc}
					\hline \hline
					& Raw          & Baseline     & Cutoff independence & Female Focus \\
					& (1)          & (2)          & (3)                 & (4)          \\
					Dep Var     & Follow Norms & Follow Norms & Gendered Language   & She Count    \\ \hline
					&              &              &                     &              \\
					After Metoo & -0.003       & -0.013*      & 0.012***            & 0.021**      \\
					& (0.005)      & (0.007)      & (0.004)             & (0.008)      \\
					&              &              &                     &              \\
					&              &              &                     &              \\
					Time FE     & No           & Yes          & Yes                 & Yes          \\
					User FE     & No           & No           & No                  & No           \\
					SE:s        & Newey- West  & Newey- West  & Newey- West         & Newey- West  \\
					& with lag=4   & with lag=4   & with lag=4          & with lag=4   \\
					Days (T)    & 319          & 287          & 287                 & 287          \\
					Data        & Year2        & Year2/1      & Year2/1             & Year2/1      \\ \hline \hline
				\end{tabular}
				\begin{tablenotes}[flushleft]
					\item[]Note: {*}{*}{*} p$<$0.01, {*}{*} p$<$0.05, {*} p$<$0.1.
				\end{tablenotes}
			\end{threeparttable}
		}
	\end{centering}	
	\caption{\label{tab:metoo-without-user-fe-autocorrelation}Metoo results without
		user fixed effects, autocorrelation taken into account}
\end{table}

\begin{table}[H]
	\begin{centering}
		\noindent\resizebox{0.7\textwidth}{!}{%
			\begin{threeparttable}
				\begin{tabular}{lcccc}
					\hline \hline
					& She               & He                & She          & He           \\
					& (1)               & (2)               & (3)          & (4)          \\
					Dep Var     & Gendered Language & Gendered Language & Follow Norms & Follow Norms \\ \hline
					&                   &                   &              &              \\
					After Metoo & 0.014***          & 0.007             & 0.023***     & -0.004**     \\
					& (0.003)           & (0.004)           & (0.005)      & (0.002)      \\
					&                   &                   &              &              \\
					Time FE     & Yes               & Yes               & Yes          & Yes          \\
					User FE     & No                & No                & No           & No           \\
					SE:s        & Newey- West       & Newey- West       & Newey- West  & Newey- West  \\
					& with lag=4        & with lag=4        & with lag=4   & with lag=4   \\
					Days (T)    & 287               & 287               & 287          & 287          \\
					Data        & Year2/1           & Year2/1           & Year2/1      & Year2/1       \\ \hline \hline
				\end{tabular}
				\begin{tablenotes}[flushleft]
					\item[]Note: {*}{*}{*} p$<$0.01, {*}{*} p$<$0.05, {*} p$<$0.1.
				\end{tablenotes}
			\end{threeparttable}
		}
	\end{centering}	
	\caption{\label{tab:metoo-subset-he-she-autocorrelation}Metoo results estimated
		separately for tweets about females and males, autocorrelation taken into account}
\end{table}

\begin{figure}[H]
	\begin{centering}
		\includegraphics[width=1\textwidth]{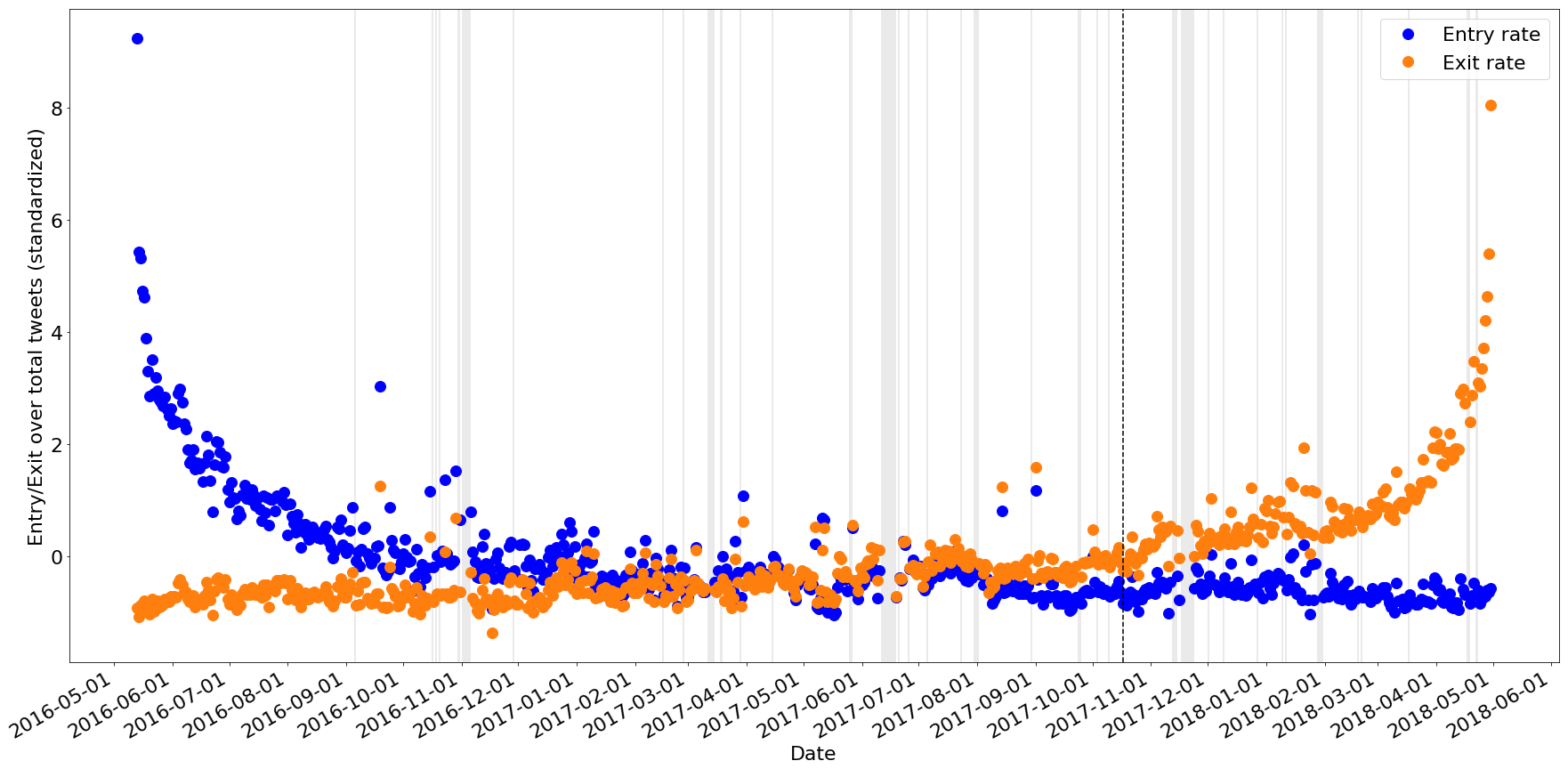}
		\par\end{centering}
	\caption{\label{fig:Entry and exit rates}Entry and exit rates}
	\begin{flushleft}
		\footnotesize{Note: The entry rate is the
			first observed tweets of users on a daily basis over total tweets.
			Instead, the exit rate is the last observed tweets. The entry and
			exit rates are standardized without any loss of pattern detection.
			The shaded areas represent missing dates. The analysis is limited
			due to unobserved tweets before and after the sample period, as well
			as for missing dates. For example, the peak of the exit rate in the
			end of the sample period is due to those dates being the last observed
			tweet for many users, but those users are likely going to tweet in
			future data which is not included. Likewise, before a region of missing
			dates the exit rate will be higher since users who's last observed
			tweet are at a missing date is counted to a date before. An alternative
			analysis would necesitate a defenition of what an active twitter user
			is.}
	\end{flushleft}
\end{figure}

\end{document}